\documentclass{article}
\usepackage{graphicx}
\usepackage{psfrag}
\usepackage{amsmath}
\usepackage{amssymb}

\title{Molecular regimes in ultracold Fermi gases} 

\author{D.S. Petrov$^{1,2}$, C. Salomon$^3$, and G.V. Shlyapnikov$^{1,4}$}
\date{}
\begin{document}
\maketitle
\noindent{\it $^1$Laboratoire de Physique Th\'eorique
et Mod\`eles Statistiques, Universit\'e Paris-Sud, 91405, Orsay Cedex,
France} \\
{\it $^2$Russian Research Center, Kurchatov Institute, Kurchatov
Square, 123182 Moscow, Russia}  \\
{\it $^3$Laboratoire Kastler Brossel, Ecole Normale Sup\'erieure,
24 rue Lhomond, 75231, Paris, France} \\
{\it $^4$Van der Waals-Zeeman Institute, University of Amsterdam, 
Valckenierstraat 65/67, 1018 XE Amsterdam, The Netherlands}

%
\begin{abstract} 

The use of Feshbach resonances for tuning the interparticle interaction
in ultracold Fermi gases has led to remarkable developments, in particular to
the creation and Bose-Einstein condensation of weakly bound diatomic molecules 
of fermionic atoms.   
These are the largest diatomic molecules obtained so far, with a size of the
order
of thousands of angstroms. They represent novel composite bosons, which exhibit
features of Fermi
statistics at short intermolecular distances. Being highly excited, these 
molecules are remarkably stable with respect to collisional
relaxation, which is a consequence of the Pauli
exclusion principle for identical fermionic atoms. The purpose of this review
is to introduce theoretical approaches and describe the physics of molecular regimes in two-component Fermi gases 
and Fermi-Fermi mixtures, focusing attention on quantum statistical effects.

\end{abstract}
%
\renewcommand{\thefootnote}{\fnsymbol{footnote}}

\tableofcontents

\section{Introduction}
\label{sec:Introduction}

\subsection{State of the art}

The field of quantum gases is rapidly 
expanding in the direction of ultracold clouds of
fermionic atoms, with the goal of revealing novel macroscopic quantum
states and achieving various regimes of superfluidity. The initial idea was to
achieve the Bardeen-Cooper-Schrieffer (BCS) superfluid phase transition
in a two-component Fermi gas,
which requires attractive interactions between the atoms of different 
components. Then, in the simplest version of this transition, at sufficiently
low temperatures fermions belonging to different components and with
opposite momenta on the Fermi surface form correlated (Cooper)
pairs in the momentum space. This leads to the appearance of a gap
in the single-particle excitation spectrum and to the phenomenon of superfluidity
(see, for example, \cite{LL}). In a dilute ultracold two-component
Fermi gas, most efficient is the formation of Cooper pairs due to
the attractive intercomponent interaction in the s-wave channel
(negative $s$-wave scattering length $a$). However, for typical
values of $a$, the superfluid transition temperature is extremely
low. For this reason, the efforts of many experimental groups have been
focused on modifying the intercomponent interaction using
Feshbach resonances. The scattering length $a$ near a Feshbach resonance can 
be tuned from $-\infty$ to $+\infty$. This has led to exciting
developments (see \cite{Varenna} for review), such as the direct observation of superfluid behavior in 
the strongly interacting regime ($n|a|^3\gtrsim 1$, where $n$ is the gas 
density) through vortex formation \cite{zw1}, 
and the study of the influence of imbalance between the two 
components of the Fermi gas on superfluidity \cite{zw2,p1,zw3,zw4,p2}. 

We focus here on the remarkable physics of weakly bound
diatomic molecules of fermionic atoms. This initially unexpected physics 
connects molecular and condensed matter physics. 
The weakly bound molecules are formed 
on the positive side of the resonance ($a > 0$) 
\cite{jila0,ens1,rudy1,jila1} and they are the largest
diatomic molecules obtained so far. Their size is of the order of
$a$ and it reaches thousands of angstroms in current experiments.
Accordingly, their binding energy is exceedingly small
($10\, \mu$K or less). Being composite bosons, these molecules
obey Bose statistics, and they have been Bose-condensed in 
experiments with $^{40}$K$_2$ \cite{jila2,jila3} at JILA and with $^6$Li$_2$
at Innsbruck \cite{rudy2,rudy3}, MIT \cite{mit1,mit2},
ENS \cite{ens2}, Rice \cite{randy2}, and Duke \cite{john}.
Nevertheless, some of the interaction properties of these
molecules reflect Fermi statistics of  the individual atoms
forming the molecule. In particular, these molecules are found
to be remarkably stable with respect to collisional decay. Being in the
highest rovibrational state, they do not undergo collisional
relaxation to deeply bound states on a time scale exceeding seconds
at densities of about $10^{13}$ cm$^{-3}$. This is more than
four orders of magnitude longer than the life time of similar
molecules consisting of bosonic atoms.  
The key idea of our discussion of homonuclear diatomic molecules formed in a two-component 
Fermi gas by atoms in different internal (hyperfine) states is to show how one obtains 
an exact universal result for the elastic
interaction between such weakly bound molecules
and how the Fermi statistics for the atoms provides a strong suppression of
their collisional relaxation into deep bound states. It is emphasized that the
repulsive character of the elastic intermolecular interaction and remarkable 
collisional stability of the molecules are the main factors allowing for their
Bose-Einstein
condensation and for prospects related to interesting manipulations with these
molecular condensates. 

Currently, a new generation of experiments is being developed
for studying degenerate mixtures of different fermionic atoms \cite{kai,walraven},
with the idea of revealing the influence of the mass difference 
on superfluid properties and finding novel types of superfluid pairing. 
On the positive side of the resonance one expects the formation 
of heteronuclear weakly bound molecules, which attracts a great deal of 
interest, in particular for creating dipolar gases. We present an analysis 
of how
the mass ratio for constituent atoms influences the elastic interaction
between the molecules and their collisional stability. The discussion is 
focused on molecules of heavy and light fermions, where one expects the 
formation of trimer bound states and the manifestation of the Efimov effect. 
We then show that a many-body system of such molecules can exhibit a 
gas-crystal quantum transition. Remarkably, the atomic system itself remains dilute, 
and the crystalline ordering is due to a relatively long-range interaction 
between the molecules originating from exchange of light fermions. 
Realization of the crystalline phase requires a very large mass ratio for 
the atoms forming a molecule in order to suppress the molecular kinetic energy. 
This can be achieved in an optical lattice for heavy atoms, where the 
crystalline phase of a dilute molecular system emerges as a superlattice,
and we discuss the related physics.

The chapter is concluded by an overview of prospects for manipulations with
the weakly bound molecules of fermionic atoms. The leading ideas 
include the achievement of ultra-low temperatures and BCS transition for atomic fermions,
creation of dipolar quantum gases, as well as observation of peculiar trimer bound states in an optical lattice. 

\subsection{Feshbach resonances and diatomic molecules}

At ultralow temperatures, when the de Broglie wavelength of atoms 
greatly exceeds the characteristic radius of interatomic interaction 
forces, atomic collisions and interactions are generally determined by 
the $s$-wave scattering. Therefore, in two-component Fermi gases one may consider 
only the interaction between atoms of different components, which 
can be tuned by using Feshbach resonances.  

The description of a many-body system near a Feshbach resonance
requires a detailed knowledge of the 2-body problem. In the vicinity of the
resonance, the energy of a colliding pair of atoms in the open channel is
close to the energy of a molecular state in another hyperfine domain (closed
channel). The coupling between these channels leads to a resonant dependence
of the scattering amplitude on the detuning $\delta$ of the closed channel
state from the threshold of the open channel, which can be controlled
by an external magnetic (or laser) field. Thus, the scattering length becomes field
dependent (see Fig.~1).

The Feshbach effect is a
two-channel problem which can be described in terms of the
Breit-Wigner scattering \cite{Breit,LL3}, and various aspects of
such problems have been discussed by Feshbach
\cite{Feshbach} and Fano \cite{Fano}. In cold atom physics the
idea of Feshbach resonances was introduced in Ref. \cite{Verhaar},
and optically induced resonances have been analyzed in Refs.
\cite{Gora,Bohn,thalhammer1,thalhammer2}.

\begin{figure}
\centerline{\includegraphics[width=11cm]{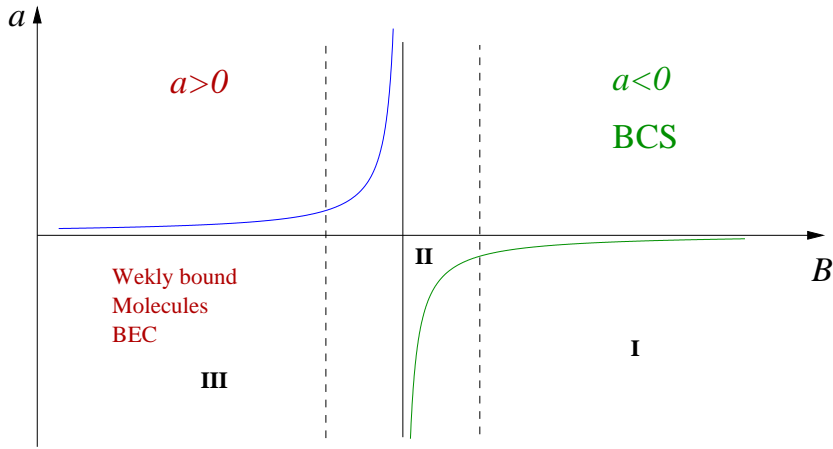}}
\caption{The dependence of the scattering length on the magnetic
field near a Feshbach resonance. The symbols I, II, and III label the regime of a weakly
interacting degenerate atomic Fermi gas, strongly interacting regime
of BCS-BEC crossover, and the
regime of weakly bound molecules. At sufficiently low temperatures region
I corresponds to the BCS superfluid pairing, and region III to
Bose-Einstein condensation of molecules.}
\end{figure}

At resonance the scattering length changes from $+\infty$ to
$-\infty$, and in the vicinity of the resonance one has the
inequality $n|a|^3\gtrsim 1$ where $n$ is the gas density. The  
gas is ten said to be in the strongly interacting regime. It is still dilute
in the sense that the mean interparticle separation greatly exceeds the
characteristic radius of the interparticle interaction $R_e$. However,
the amplitude of binary interactions (scattering length) is larger
than the mean separation between particles, and in the quantum 
degenerate regime the conventional mean field approach is no longer valid.

For large detuning from resonance the gas is in the weakly
interacting regime, i.e. the inequality $n|a|^3\ll 1$ is
satisfied. On the negative side of the resonance ($a<0$), at sufficiently 
low temperatures of the two-species Fermi gas one expects the BCS pairing
between distinguishable fermions, well described in literature
\cite{LL}. On the positive side ($a>0$) two fermions belonging to
different components form diatomic molecules. For $a\gg R_e$
these molecules are weakly bound and their size is of the order of $a$. 

The crossover from the BCS to BEC behavior has recently attracted now a great deal
of interest, in particular with respect to the nature of
superfluid pairing, transition temperature, and elementary
excitations. This type of crossover has
been earlier discussed in the literature in the context of superconductivity
\cite{Eag,Leg,Noz,Rand} and in relation to superfluidity in
two-dimensional films of $^3$He \cite{M,MYu}.
The idea of resonant coupling through a Feshbach resonance for
achieving a superfluid phase transition in ultracold two-component
Fermi gases has been proposed in Refs. \cite{Holland1,Tim}, and for the two-dimensional
case it has been discussed in Ref.~\cite{BPS}.

The two-body physics of the Feshbach resonance is the most transparent 
if the (small) background scattering length is neglected. Then for low
collision
energies $\varepsilon$ the scattering amplitude is given by
\cite{LL3} :
\begin{equation} \label{FE}
F(\varepsilon)=-\frac{\hbar\gamma/\sqrt{2\mu}}
{\varepsilon+\delta+i\gamma\sqrt{\varepsilon}},
\end{equation}
where the quantity $\hbar\gamma/\sqrt{2\mu}\equiv W$ characterizes
the coupling between the open and closed channels and $\mu$ is the
reduced mass of the
two atoms. The scattering length is $a=-F(0)$. In Eq.~(\ref{FE}) the
detuning
$\delta$ is positive if the bound molecular state is below the
continuum of the
colliding atoms. Then for $\delta>0$ the scattering
length is positive, and for $\delta<0$ it is negative. Introducing
a characteristic length
\begin{equation} \label{RW}
R^*=\hbar^2/2\mu W
\end{equation}
and expressing the scattering amplitude through  the relative
momentum of particles $k=\sqrt{2\mu\varepsilon}/\hbar$, we can rewrite Eq.~(\ref{FE})
in the form:
\begin{equation} \label{Fk}
F(k)=-\frac{1}{a^{-1}+R^*k^2+ik}.
\end{equation}
The validity of Eq.~(\ref{Fk}) does not require the condition
$kR^*\ll 1$. At the same time, this equation formally coincides
with the amplitude of scattering of slow particles by a potential
with the same scattering length $a$ and an effective range
$R=-2R^*$, obtained under the condition $k|R|\ll 1$.

The length $R^*$ is an intrinsic parameter of a Feshbach
resonance. It characterizes the width of the resonance.
From Eqs.~(\ref{FE}) and (\ref{RW}) we see that small $W$ and,
consequently, large $R^*$ correspond to narrow resonances, whereas large
$W$ and small $R^*$ lead to wide resonances. The term ``wide'' is
generally used when the length $R^*$ drops out of the problem,
which according to Eq.~(\ref{Fk}) requires the condition $kR^*\ll 1$.
In a quantum degenerate atomic Fermi gas the characteristic momentum
of particles is the Fermi momentum $k_F=(3\pi^2n)^{1/3}$. 
Thus, in the strongly interacting regime and on the negative side of 
the resonance ($a<0$), for a given $R^*$ the condition of the wide 
resonance depends on the gas density $n$ and takes the form $k_FR^*\ll 1$
\cite{Bruun,Bruun2,Palo,Eric,Ho3}. 

For $a>0$ one has weakly bound 
molecular states (it is certainly assumed that the characteristic
radius of interaction $R_e\ll a$), and for such molecular systems 
the criterion of the wide resonance is different \cite{Petrovbosons,PSS}. 
The binding energy of the weakly bound
molecule state  is determined by the pole  of
the scattering amplitude (\ref{Fk}). One then finds
\cite{Petrovbosons,PSS} that this state exists only for $a>0$, and
under the condition
\begin{equation} \label{Ra}
R^*\ll a
\end{equation}
the binding energy is given by
\begin{equation}  \label{binding}
\varepsilon_0=\hbar^2/2\mu a^2.
\end{equation}
The wavefunction of such weakly bound molecular state has only
a small admixture of the closed channel, and the size of the molecule
is $\sim a$. The characteristic momenta of the atoms in the molecule are
of the order of $a^{-1}$ and in this respect the inequality (\ref{Ra}) 
represents the criterion of a wide resonance for the molecular system.

Under these conditions atom-molecule and molecule-molecule interactions
are determined by a single parameter -- the atom-atom scattering
length $a$. In this sense, the problem becomes universal. It is
equivalent to the interaction problem for the two-body potential
which is characterized by a large positive scattering length $a$
and has a potential well with a weakly bound molecular state. The
picture remains the same when the background scattering length
can not be neglected, although the condition of a wide
resonance can be somewhat modified \cite{Karen}.

Most ongoing experiments with Fermi gases of atoms in two different 
internal (hyperfine) states use wide Feshbach resonances \cite{randy1}. 
For example, weakly bound molecules
$^6$Li$_2$ and $^{40}$K$_2$ have been produced in experiments
\cite{ens1,rudy1,jila1,jila2,jila3,mit1,mit2,rudy2,rudy3,ens2,randy2}
by using Feshbach resonances with a length $R^*\lesssim 20$\AA, and
for the achieved values of the
scattering length $a$ (from $500$ to $2000$\AA) the ratio $R^*/a$
was smaller than $0.1$. In this review we will consider 
the case of a wide Feshbach resonance.

\section{Homonuclear diatomic molecules in Fermi gases}

\subsection{Weakly interacting gas of bosonic molecules. Molecule-molecule
elastic interaction}  

As we have shown in the previous section, the size of weakly bound bosonic
molecules 
formed at a positive atom-atom scattering length $a$ in a two-species Fermi gas 
(region III in Fig.~1) is of the order of $a$. Therefore, at densities
such that $na^3\ll 1$, the atoms form a weakly interacting gas of these
molecules.
Moreover, under this condition at temperatures sufficiently lower than the
molecular binding energy 
$\varepsilon_0$ and for equal concentrations of the two 
atomic components, practically all atoms are converted into
molecules \cite{Servaas}. This is definitely the case at temperatures 
below the temperature of quantum degeneracy $T_d=2\pi\hbar^2n^{2/3}/M$
(the lowest one in the case of fermionic atoms with different masses, with $M$
being the mass of the heaviest atom).
One can clearly see this by comparing $T_d$ with $\varepsilon_0$ given by Eq
(\ref{binding}).
Thus, one has a weakly interacting molecular Bose gas and the first
question is related to the elastic interaction between the
molecules. 

For a weakly interacting gas the interaction energy in the system 
is equal to the sum of pair interactions and the energy per particle
is $ng$ ($2ng$ for a non-condensed Bose gas),
with $g$ being the coupling constant. In our case this coupling constant
is given by $g=4\pi\hbar^2a_{dd}/(M+m)$, where $a_{dd}$ is the scattering 
length for the molecule-molecule (dimer-dimer) elastic $s$-wave scattering,
and $M$, $m$ are the masses of heavy and light atoms, respectively. 
The value of $a_{dd}$ is important for evaporative cooling of the
molecular gas to the regime of Bose-Einstein condensation and for
the stability of the condensate. The Bose-Einstein condensate is stable
for repulsive intermolecular interaction ($a_{dd}>0$), and for $a_{dd}<0$
it collapses.

We thus see that for analyzing macroscopic properties of the molecular
Bose gas one should first solve the problem of elastic interaction (scattering)
between two molecules. In this section we present the exact solution of
this problem for homonuclear molecules formed by fermionic atoms of different
components (different internal states) in a two-component Fermi gas. The case of
$M\neq m$ will be discussed in Section 3. The solution for $M=m$ was obtained in 
Refs.~\cite{Petrov1,PSS} assuming that the atom-atom scattering length $a$ 
greatly exceeds the characteristic radius of interatomic potential:
\begin{equation}   \label{aRe}
a\gg R_e.
\end{equation}
Then, as in the case of the 3-body problem with fermions
\cite{STM,Danilov,efimov,Petrov2}, the amplitude of elastic
interaction is determined only by $a$ and can be found in the zero-range
approximation for the interatomic potential.

This approach was introduced in the two-body physics by Bethe and
Peierls \cite{Bethe}. The leading idea is to solve the
equation for the free relative motion of two particles placing a
boundary condition on the wavefunction $\psi$ at a vanishing
interparticle distance $r$:
\begin{equation}\label{twobody.Bethe-Reierls}
\frac{(r\psi)'}{r\psi}=-\frac{1}{a},\,\,\,\,\,\,\,r\rightarrow 0,
\end{equation}
which can also be rewritten as
\begin{equation}\label{boundary1}
\psi\propto (1/r-1/a),\,\,\,\,\,\,\,r\rightarrow 0.
\end{equation}
One then gets the correct expression for the wavefunction at
distances $r\gg R_e$. when $a\gg R_e$,
Eq.~(\ref{boundary1}) correctly describes the
wavefunction of weakly bound and continuum states even at
distances much smaller than $a$. 

\begin{figure}
\label{fig6}
\centerline{\includegraphics[width=8cm]{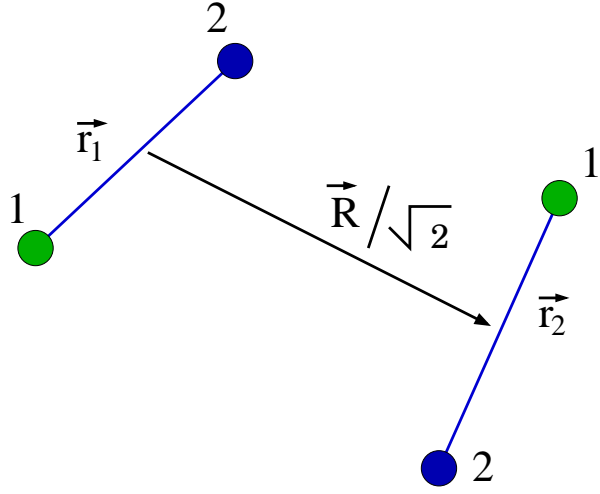}}
\caption{Set of coordinates for the four-body problem.}
\end{figure}

We now use the Bethe-Peierls approach for the problem of 
elastic molecule-molecule (dimer-dimer) scattering which is a 4-body 
problem described by the Schr\"odinger equation
\begin{eqnarray}\label{4bodySchrU}
\!\!\Big\{\!\!-\frac{\hbar^2}{m}(\nabla_{{\bf r}_1}^2\!\!-\!\nabla_{{\bf
r}_2}^2\!\!-\!\nabla_{\bf
{R}}^2)\!+\!U(r_1)\!+\!U(r_2)\!
+\!\!\sum_{\pm}U[({\bf r}_1\!+\!{\bf r}_2\!\pm\!\sqrt{2}{\bf
R})/2]\!-\!E\Big\}\Psi\!=\!0,\!
\end{eqnarray}
where $m$ is the atom mass. Labeling fermionic atoms in different internal
states
by the symbols $\uparrow$ and $\downarrow$, the distance between two given
$\uparrow$ and $\downarrow$ fermions is ${\bf r}_1$, and ${\bf r}_2$ is the
distance 
between the other two. The distance between the centers of mass of these pairs 
is ${\bf {R}}/\sqrt{2}$, and $({\bf r}_1+{\bf r}_2
\pm\sqrt{2}{\bf R})/2$ are the separations between $\uparrow$ and $\downarrow$
fermions in the other two possible $\uparrow\downarrow$ pairs (see Fig.~2). The
total energy is
$E=-2\varepsilon_0+\varepsilon$, with
$\varepsilon$ being the collision energy, and
$\varepsilon_0=-\hbar^2/ma^2$ the binding
energy of a dimer.
The wavefunction $\Psi$ is symmetric with respect to the
permutation of bosonic $\uparrow\downarrow$ pairs and antisymmetric with
respect to
permutations of identical fermions:
\begin{equation}\label{symmetry}
\!\!\Psi({\bf r}_{1},{\bf r}_{2},{\bf {R}})\!\!=\!
\Psi({\bf r}_{2},{\bf r}_{1},\!-{\bf {R}})\!\! 
=\!-\Psi\!\!\left(\!\frac{{\bf r}_1\!+\!{\bf r}_2\! \pm\!\sqrt{2}{\bf R}}{2},\,
\frac{{\bf r}_1\!+\!{\bf
r}_2 \!\mp\!\sqrt{2}{\bf R}}{2}, \pm\frac{{\bf r}_1\!-\!{\bf r}_2} {\sqrt{2}}\right)\!\!.\!\!\!\!\!\!\!\!
\end{equation}

For the weak binding of atoms in the molecule assuming 
that the 2-body scattering length satisfies the inequality (\ref{aRe}), 
at all interatomic distances (even much smaller than $a$) except for very 
short separations of the order of or smaller than $R_e$, the motion of atoms in
the
4-body system is described by the free-particle Schr\"odinger equation
\begin{equation}\label{4bodySchr}
-\left[\nabla_{{\bf r}_1}^2+\nabla_{{\bf r}_2}^2+\nabla_{\bf
{R}}^2+\frac{mE}{\hbar^2}\right]\Psi=0.
\end{equation}
The correct description of this motion requires the 4-body wavefunction $\Psi$ 
to satisfy the Bethe-Peierls boundary
condition for the vanishing distance in any pair of $\uparrow$ and $\downarrow$
fermions, i.e. for ${\bf r}_1\rightarrow 0$, ${\bf r}_2\rightarrow 0$, and
${\bf
r}_1+{\bf r}_2 \pm\sqrt{2}{\bf R}\rightarrow 0$. Due to the symmetry condition 
(\ref{symmetry}) it is necessary
to require a proper behavior of $\Psi$ only at one of these boundaries. For
${\bf
r}_1\rightarrow 0$ the boundary condition reads:
\begin{equation}\label{boundary}
\Psi({\bf r}_1,{\bf r}_2,{\bf {R}})\rightarrow f({\bf r}_2,{\bf
R})(1/4\pi r_1\,-1/4\pi a).
\end{equation}
The function $f({\bf r}_2,{\bf R})$ contains the information about the second
pair of
particles when the first two are on top of
each other.

In the ultracold limit, where  
\begin{equation}  \label{ka}
ka\ll 1,
\end{equation}
the molecule-molecule scattering is dominated by the
contribution of the $s$-wave channel. The inequality (\ref{ka}) is equivalent
to
$\varepsilon\ll \varepsilon_0$ and, hence, the $s$-wave scattering  can be
analyzed
from the solution of
Eq.~(\ref{4bodySchr}) with $E=-2\varepsilon_0<0$. For large $R$ the
corresponding
wavefunction takes the form
\begin{equation}   \label{asymptote0}
\Psi\approx\phi_0(r_1)\phi_0(r_2)(1-\sqrt{2}a_{dd}/R);\,\,\,\,\,R\gg a,
\end{equation}
where the wavefunction of a weakly bound molecule is given by
\begin{equation}  \label{phi0}
\phi_0(r)=\frac{1}{\sqrt{2\pi a}\,r}\exp(-r/a).
\end{equation}  
Combining Eqs.~(\ref{boundary}) and (\ref{asymptote0}) we obtain the asymptotic
expression for 
$f$ at large distances $R$:
\begin{equation}\label{dimerdimer.swave}
f({\bf r}_2,{\bf R})\approx (2/r_2a)\exp{(-r_2/a)}(1-\sqrt{2}a_{dd}/R);\,\,\
\,R\gg a.
\end{equation}

In the case of  $s$-wave scattering the function $f$ depends only on
three variables: the absolute values of ${\bf r}_2$ and ${\bf {R}}$, and the
angle
between them. We now derive and solve the equation for $f$. The value of the
molecule-molecule scattering length $a_{dd}$ is
then deduced from the behavior of $f$ at large $R$ determined by
Eq.~(\ref{dimerdimer.swave}).

We first establish a general form of the wavefunction $\Psi$ satisfying
Eq.~(\ref{4bodySchr}), with the boundary condition (\ref{boundary}) and
symmetry
relations (\ref{symmetry}). In our case the total energy $E=-2\hbar^2/ma^2<0$,
and the Green
function of Eq.~(\ref{4bodySchr}) reads
\begin{equation} \label{Green4body}
G(X)=(2\pi)^{-9/2}(Xa/\sqrt{2})^{-7/2} K_{7/2}(\sqrt{2}\,X/a),
\end{equation}
where $X=|S-S'|$, $K_{7/2}(\sqrt{2}X/a)$ is the deacaying Bessel function, and $S=\{{\bf r}_1,{\bf r}_2,{\bf R}\}$ is a 9-component
vector.
Accordingly, 
$|S-S'|=\sqrt{({\bf r}_1-{\bf r'}_1)^2+({\bf r}_2-{\bf r'}_2)^2+({\bf R}-{\bf
R'})^2}$. The 4-body wavefunction $\Psi$ is regular everywhere except for
vanishing
distances between $\uparrow$ and $\downarrow$ fermions. Therefore, it can be
expressed through $G(|S-S'|)$ with coordinates $S'$ corresponding to a
vanishing
distance between $\uparrow$ and $\downarrow$ fermions, i.e. for 
${\bf r'}_1\rightarrow 0$, ${\bf r'}_2\rightarrow 0$, and $({\bf r'}_1+{\bf
r'}_2\pm\sqrt{2}{\bf R'})/2\rightarrow 0$. Thus, for the wavefunction $\Psi$
satisfying the
symmetry relations (\ref{symmetry}) we have
\begin{eqnarray}\label{Psi4body}
&&\Psi(S)=\Psi_0+\int d^3r'd^3R'\Big[ G(|S-S_1|)+G(|S-S_2|) \nonumber \\ 
&&-G(|S-S_+|)-G(|S-S_-|)\Big]h({\bf r'},{\bf R'}),
\end{eqnarray}
where $S_1=\{0,{\bf r}',{\bf {R}}'\},\;S_2=\{{\bf r}',0,-{\bf {R}}'\}$, and
$S_\pm=\{{\bf r}'/2\pm{\bf {R}}'/\sqrt{2},{\bf r}'/2\mp{\bf {R}}'/\sqrt{2}
\mp{\bf
r}'\sqrt{2}\}$.
The function $\Psi_0$ is a properly symmetrized finite solution of
Eq.~(\ref{4bodySchr}),
regular at any distances between the atoms. For
$E<0$, non-trivial solutions of this type do not exist and we have to put
$\Psi_0=0$. The function $h({\bf r}_2,{\bf R})$ has to be determined by
comparing
$\Psi$ in Eq.~(\ref{Psi4body}) at ${\bf r}_1\rightarrow 0$, with the boundary
condition
(\ref{boundary}).   

Considering the limit ${\bf r}_1\rightarrow 0$ we extract the leading
terms on the right hand side of Eq.~(\ref{Psi4body}). These are the terms that
behave
as $1/r_1$
or remain finite in this limit. The last three terms in the square brackets in
Eq.~(\ref{Psi4body}) provide a finite contribution
\begin{equation} \label{reg1}
\int d^3r'd^3R'\,h({\bf r}',{\bf R}')\Big[G(|{\bar S}_2-S_2|)
-G(|{\bar S}_2-S_+|)-G(|{\bar S}_2-S_-|)\Big],
\end{equation}    
where ${\bar S}_2=\{0,{\bf r}_2,{\bf R}\}$. To find the  contribution of
the
first term in the square brackets, we subtract and add an
auxiliary quantity
\begin{equation} \label{aux0}
h({\bf r}_2,{\bf R})\int G(|S-S_1|)d^3r'd^3R'
=\frac{h({\bf r}_2,{\bf R})}{4\pi r_1}\exp{(-\sqrt{2}r_1/a)}.
\end{equation}
The result of the subtraction yields a finite contribution which for
$r_1\rightarrow
0$ can be written as
\begin{eqnarray} \label{reg2}
&& \int d^3r'd^3R'[h({\bf r}',{\bf R}')-h({\bf r}_2,{\bf R})]G(|S-S_1|)  
\nonumber \\ 
&& =P \int d^3r'd^3R'[h({\bf r}',{\bf R}')-h({\bf r}_2,{\bf R})]G(|{\bar
S}_2-S_1|);\,\,\,\,\,\,r_1\rightarrow 0,
\end{eqnarray}
with the symbol $P$ denoting the principal value of the integral over $dr'$
(or
$dR'$). A detailed derivation of Eq.~ (\ref{reg2}) and the proof that the
integral in
the second line of this equation is convergent are given in Ref. \cite{PSS}. 

In the limit $r_1\rightarrow 0$, the right hand side of Eq.~(\ref{aux0}) is
equal to
\begin{equation} \label{aux}
h({\bf r}_2,{\bf R})(1/4\pi r_1-\sqrt{2}/4\pi a).
\end{equation}
We thus  find that for ${\bf r}_1\rightarrow 0$ the wavefunction $\Psi$ of
Eq.~(\ref{Psi4body}) takes the form
\begin{equation} \label{Psilim}
\Psi({\bf r}_1,{\bf r}_2,{\bf R})=\frac{h({\bf r}_2,{\bf R})}{4\pi r_1}+{\cal
R};\,\,\,\,\,\,{\bf r}_1\rightarrow 0,
\end{equation}
where ${\cal R}$ is the sum of regular $r_1$-independent terms given by
Eqs.~(\ref{reg1}) and (\ref{reg2}), and by the second term on the right hand
side of
Eq.~(\ref{aux}). Equation (\ref{Psilim}) must coincide
with Eq.~(\ref{boundary}), and comparing the singular terms of these equations
we
find $h({\bf r}_2,{\bf R})=f({\bf r}_2,{\bf R})$. As  the quantity ${\cal R}$
must
coincide with the regular term of Eq.~(\ref{boundary}), equal to $-f({\bf r}_2
{\bf
R})/4\pi a$, we obtain the following equation for the function $f$:
\begin{eqnarray}\label{main}
&&\int d^3r'd^3R'\Big\{G(|{\bar S}-S_1|)[f({\bf r}',{\bf {R}}')-f({\bf r},{\bf
{R}})]
+\bigl[G(|{\bar S}-S_2|)\nonumber \\
&&-\sum_\pm G(|{\bar S}-S_\pm|)\bigr]f({\bf r}',{\bf {R}}')\Big\} 
=(\sqrt{2}-1)f({\bf r},{\bf {R}})/4\pi a.
\end{eqnarray}
Here ${\bar S}=\{0,{\bf r},{\bf {R}}\}$, and we omitted the symbol of the principal
value
for the integral in the first line of Eq.~(\ref{main}). 

As we have already mentioned above, for $s$-wave scattering the function $f({\bf
r},{\bf R})$ depends only on the absolute values of ${\bf r}$ and ${\bf R}$ and
on the angle between them. Thus, Eq.~(\ref{main}) is an integral equation for the
function of three variables. In order to find the molecule-molecule scattering length, it 
is more convenient to transform Eq.~(\ref{main}) into an equation for the momentum-space function $f({\bf k},{\bf
p})=\int d^3rd^3Rf({\bf r},{\bf R})\exp(i{\bf k\cdot r}/a+i{\bf p\cdot R}/\sqrt{2}a)$, which yields the following expression:
\begin{eqnarray}\label{momentum1}
\!\!\!\!\!&\!\!&\sum_\pm\!\!\int\!\!
\frac{f({\bf k}\pm ({\bf p}'-{\bf p})/2,{\bf p}')\,{\rm
d}^3p'}{2+\!p'^2/2+\!({\bf k}\pm ({\bf p}'\!-{\bf
p})/2)^2+({\bf k}
\pm ({\bf p}'\!+{\bf p})/2)^2}\nonumber\\
\!\!\!\!&\!\!&\!=\!\int\!\! \frac{f({\bf k}',-{\bf p})\,{\rm
d}^3k'}{2\! +\! k'^2\!
+\! k^2\! +\! p^2/2}- \frac{2\pi^2(1 +\! k^2\! +p^2/2)f({\bf
k},{\bf
p})}{\sqrt{2 + k^2 +p^2/2}+ 1}.
\end{eqnarray} 
By making the substitution $f({\bf k},{\bf p})=(\delta({\bf p})+g({\bf
k},{\bf
p})/p^2)/(1+k^2)$ we reduce Eq.~(\ref{momentum1}) to an inhomogeneous
equation for
the function $g({\bf k},{\bf p})$: 
\begin{eqnarray}         \label{g1}
\!\!\!\!\!\!\!\!\!\!&&\frac{1}{(1+k^2+p^2/4)^2-({\bf k}{\bf p})^2}+\frac{2\pi^2(1+k^2+p^2/2)g({\bf k}, {\bf p})}{p^2(1+k^2)(\sqrt{2+k^2+p^2/2}+1)} \nonumber \\
\!\!\!\!\!\!\!\!\!\!&&\!=\!\!-\!\!\sum_{\pm}\!\!\int\!\! \frac{g({\bf k}\pm({\bf p}'-{\bf p})/2, {\bf p}')d^3p'}{p'^2(2\!+\!p'^2/2+\!({\bf k}\pm\! ({\bf p}'\!\!\!-\!\!{\bf
p})/2)^2\!\!+\!\!({\bf k}\!\pm \!({\bf p}'\!\!+\!\!{\bf p})/2)^2)(1\!\!+\!\!({\bf k}\!\pm ({\bf p}'\!\!-\!\!{\bf p})/2)^2)} \nonumber \\ 
\!\!\!\!\!\!\!\!\!\!&&+\int \frac{g({\bf k}',-{\bf p})d^3k'}{p^2(2+k'^2+k^2+p^2/2)(1+k'^2)}.
\end{eqnarray}
In the case of $s$-wave scattering the function $g({\bf k},{\bf p})$ depends 
on the absolute values of ${\bf k}$ and ${\bf p}$ and on the angle between these vectors.       
For ${\bf p}\rightarrow 0$ this function tends to a finite value independent of ${\bf k}$. 
As one can easily establish on the basis of Eq.~(\ref{dimerdimer.swave}) and the definition of $g({\bf k},{\bf p})$,
the molecule-molecule scattering length is given by
$a_{dd}=-2\pi^2a\lim_{{\bf p}\rightarrow 0}g({\bf k},{\bf p})$. Numerical calculations from 
Eq.~(\ref{g1}) give with 2\% accuracy \cite{PSSJ}:
\begin{equation}   \label{add}
a_{dd}=0.6 a>0.
\end{equation}
This result was first obtained in Refs.~\cite{Petrov1,PSS} from the direct numerical solution
of Eq.~(\ref{main}) by fitting the obtained $f({\bf r}, {\bf R})$ with the asymptotic form
(\ref{dimerdimer.swave}) at large $R$. The calculations show the absence of 4-body weakly bound states, and the
behavior of $f$ at small $R$ suggests a soft-core repulsion between dimers, with a range $\sim a$. 

The result of Eq.~(\ref{add}) is exact, and it indicates the
stability of
molecular BEC with respect to collapse. Compared to earlier studies which
assumed
$a_{dd}=2a$ \cite{Rand,strinati}, Eq.~(\ref{add}) gives almost twice as small a
sound
velocity of the molecular condensate and a rate of elastic collisions smaller
by an
order of magnitude. The result of Eq.~(\ref{add}) has been confirmed by Monte
Carlo calculations 
\cite{giorgini} and by calculations within the diagrammatic approach
\cite{maxim,gurarie}. An approximate diagrammatic approach leading to $a=0.75a$
has been developed in Ref.~\cite{strinati}.

\subsection{Suppression of collisional relaxation}

Weakly bound dimers that we are considering are diatomic molecules in the 
highest ro-vibrational state (see Fig.~3). They can undergo relaxation
into deeply bound
states in their collisions with each other: for example, one of the colliding
molecules may relax to
a deeply bound state while the other one dissociates \cite{comment1}. The
released energy is the binding
energy of the final deep state, which is of the order of $\hbar^2/mR_e^2$. It is
transformed 
into the kinetic energy of the particles in the outgoing collision channel and
they escape from the
trapped sample. Therefore, the process of collisional relaxation of weakly
bound
molecules determines  the lifetime of a gas of these molecules and
possibilities to Bose-condense this gas.

\begin{figure}
\label{fig7}
\centerline{\includegraphics[width=8cm]{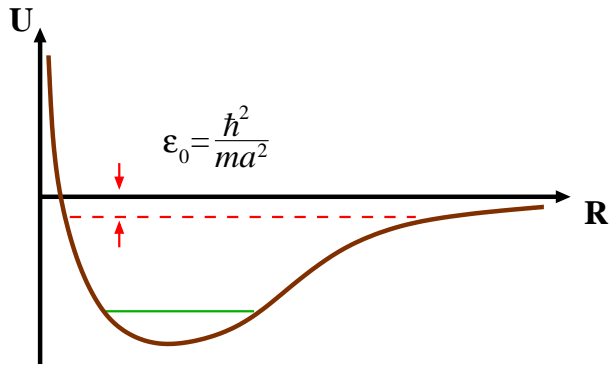}}
\caption{Interaction potential $U$ as a function of the distance $R$
between two distinguishable fermionic atoms. The dashed line shows the
energy level of the weakly bound molecule, and the solid line the energy level
of
a deeply bound state.}
\end{figure} 

We now show that collisional relaxation is suppressed due to {\it Fermi 
statistics} for atoms in combination with a {\it large size} of weakly bound
molecules \cite{Petrov1,PSS}. The binding energy of the molecules is 
$\varepsilon_0=\hbar^2/ma^2$ and their size is $\sim a\gg R_e$. The
size of deeply bound states is of the order of $R_e$. Therefore, the relaxation
process may occur when at least three
fermionic atoms are at distances $\sim R_e$ with respect to each other. As two of them are
necessarily
identical, due to the Pauli exclusion principle the relaxation probability
acquires
a small factor proportional to a power of $(qR_e)$, where $q\sim 1/a$ is a
characteristic momentum of the atoms in the weakly bound molecular state. 

Relying on the inequality $a\gg R_e$ we outline a method that allows one to
establish the
dependence of the relaxation rate on the scattering length $a$, without going
into a
detailed analysis of the short-range behavior of
the system. It is assumed that the amplitude of the inelastic
relaxation process is much smaller than the amplitude of elastic scattering.
Then
the dependence of the relaxation rate on $a$ is related only to the
$a$-dependence
of the initial-state 4-body wavefunction $\Psi$. We again consider the
ultracold limit 
described by the condition (\ref{ka}), where the relaxation process is
dominated by the 
contribution of the $s$-wave molecule-molecule scattering. 

\begin{figure}
\label{fig8}
\centerline{\includegraphics[width=8cm]{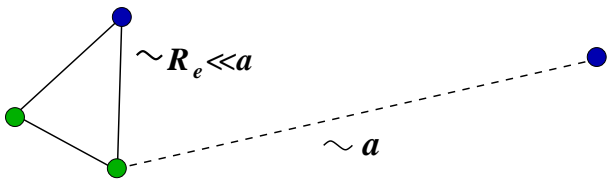}}
\caption{Configuration space contributing to the relaxation probability.}
\end{figure} 

The key point is that the relaxation process requires only three atoms to
approach each other to
short distances of the order of $R_e$. The fourth particle can be far away 
from these three and, in this respect, does not participate in the
relaxation process. This distance is of the order of the size of a molecule,
which is
$\sim a\gg R_e$. 
We thus see that the configuration space contributing to the relaxation
probability can be viewed as a
system of three atoms at short distances $\sim R_e$ from each other and a 
fourth
atom separated from this system by a large distance $\sim a$ (see Fig.~4). In
this case the
4-body wavefunction decomposes into a product: 
\begin{equation}\label{decomp}
\Psi=\eta({\bf z})\Psi^{(3)}(\rho,\Omega),
\end{equation}
where  $\Psi^{(3)}$ is the wavefunction of the 3-fermion system, $\rho$ and
$\Omega$
are the hyperradius and the set of hyperangles for these fermions,  ${\bf z}$
is the
distance between their center of mass and the fourth atom. The wavefunction
$\eta({\bf z})$ describes the motion of this atom and is normalized to unity. 
Note that Eq.~(\ref{decomp}) remains valid for any hyperradius $\rho\ll |{\bf
z}|\sim a$. 

The transition to a deeply bound 2-body state occurs in the system of three atoms
and
does not change the wavefunction of the fourth atom, $\eta({\bf z})$.
Therefore,
averaging the transition probability over the motion of the fourth particle,
the
rate constant for inelastic relaxation in dimer-dimer collisions can be written as  
\begin{equation} \label{alpha34}
\alpha_{rel}=\alpha^{(3)}\int |\eta({\bf z})|^2d^3z=\alpha^{(3)},
\end{equation}
where $\alpha^{(3)}$ is the rate constant for relaxation in the 3-atom system.

At interatomic distances $\sim R_e$ where the relaxation occurs, as well as at
all 
distances where the hyperradius $\rho\ll a$,  the
wavefunction $\Psi^{(3)}$ is determined by the Schr\"odinger equation
with zero energy and, hence, depends on the scattering
length $a$ only through a normalization coefficient: 
\begin{equation} \label{Psipsi}
\Psi^{(3)}=A(a)\psi;\,\,\,\,\,\,\,\,\,\rho\ll a,
\end{equation}
where the function $\psi$ is independent of $a$. 
The probability of relaxation and, hence, the relaxation rate constant 
are proportional to $|\Psi|^2$ at distances $\sim R_e$. We thus  have
\begin{equation} \label{alphaA}
\alpha_{rel}=\alpha^{(3)}\propto |A(a)|^2.
\end{equation}
The goal then is to find the coefficient $A(a)$, which determines the
dependence of
the relaxation rate on $a$. 

For this purpose it is sufficient to consider distances where $a\gg\rho\gg R_e$ 
and Eq.~(\ref{Psipsi}) is still valid. Then, using the zero-range approximation
we find the coordinate dependence of the three-body wavefunction $\Psi^{(3)}$. 
The derivation is presented in Ref.~\cite{PSS} and the result is:
\begin{eqnarray} 
\Psi^{(3)}=A(a)\Phi_{\nu}(\Omega)\rho^{\nu-1},\,\,\,\,\,\rho\ll a,
\label{Psi3nu}
\end{eqnarray}
where $\Phi_{\nu}(\Omega)$ is a normalized function of hyperangles, and 
the coefficient $\nu$ depends on the symmetry of $\Psi^{(3)}$. 
The $a$-dependence of the prefactor $A(a)$ can be determined from
the following scaling arguments for the 4-body problem. The scattering length
$a$ 
is the only length scale in our problem and we can measure all distances in 
units of $a$. Using two rescaled coordinates, 
$\rho=a\rho'$ and ${\bf z}=a{\bf z}'$, we see that $\Psi^{(3)}$ in
Eqs.~(\ref{Psi3nu}) and (\ref{decomp})
becomes a function of $\rho/a$, multiplied by $A(a)a^{\nu-1}$. The wavefunction
$\eta({\bf z})$ is normalized to unity and hence it is a function of ${\bf
z}/a$,
multiplied by $a^{-3/2}$. Accordingly, the 4-body wavefunction $\Psi$ of 
Eq.~(\ref{decomp}) is a function of rescaled coordinates, multiplied by
the coefficient 
$A(a)a^{\nu-5/2}$. By applying the same rescaling to Eq.~(\ref{asymptote0}) and
using Eq.~(\ref{phi0}) we see that the same coefficient should be proportional to
$a^{-3}$. 
Therefore, $A(a)\propto a^{-\nu-1/2}$ and $\alpha_{rel}\propto a^{-s}$, where
$s=2\nu+1$.

The strongest relaxation channel corresponds to the lowest value of $\nu$.
It is achieved in the case of $p$-wave symmetry in the three-body system
described by the 
wavefunction $\Psi^{(3)}$ and is equal to $\nu=0.773$, which leads to $s=2.55$.
Assuming that the short-range physics is characterized by the length scale 
$R_e$ and the energy scale $\hbar^2/mR_e^2$ we can restore the dimensions and
write:
\begin{equation} \label{alphadd}
\alpha_{rel}=C(\hbar R_e/m)(R_e/a)^s;\,\,\,\,\,\,\,\,\,s=2.55,
\end{equation}
where the coefficient $C$ depends on a particular system and can not be
obtained
using the zero-range approximation. 

Note that the $p$-wave symmetry in the three-body system corresponds to 
$p$-wave scattering
of a fermionic atom of one of the molecules (referred to as the 3-rd fermion)
on the other molecule. 
Then the 4-th particle also undergoes $p$-wave scattering on this molecule 
in such a way that the total orbital angular momentum of the molecule-molecule
collision is 
equal to zero. Since the 3-rd and 4-th fermions are bound to each other in the
molecular state with a size 
$\sim a$, the relative momentum of their collisions with the other molecule is
$\sim 1/a$ and
such $p$-wave collisions are not at all suppressed. The relaxation channel
corresponding 
to $s$-wave scattering of the 3-rd fermion on the molecule leads to 
$\nu=1.1662$ and hence to the relaxation rate proportional to $a^{-3.33}$ as in
the case of ultracold
atom-molecule collisions \cite{Petrov1,PSS}.
Thus, for large $a$ this mechanism can be omitted.  
The channels, where the 3-rd fermion (and the 4-th one) 
scatters on the molecule with orbital angular momentum $l>1$, lead to even
stronger decrease of the relaxation rate with increasing $a$ and hence can
be neglected.  

\subsection{Collisional stability and molecular BEC}

Equation (\ref{alphadd}) implies a remarkable collisional stability of weakly
bound molecules 
consisting of fermionic atoms in two different internal states and a
counterintuitive
decrease of the relaxation rate with increasing the two-body scattering length
$a$. 
For currently achieved values of the scattering length $a\!\sim\!1000$ \AA,
the
suppression factor $(R_e/a)^s$ for the relaxation process is about 4 orders of
magnitude.
This effect is due to {\it Fermi statistics} for the atoms. It is not present
for weakly 
bound molecules of bosonic atoms, even if they have the same large size.
Indeed, as the size
of weakly bound molecules is $\sim a$, identical fermionic atoms participating
in the 
relaxation process have very small relative momenta $k\sim 1/a$. Hence, the
probability
that they approach each other to short distances $\sim R_e$ where the
relaxation
transitions occur, should be suppressed as $(kR_e)^2\sim (R_e/a)^2$
compared to the case of molecules of bosonic atoms. The exponent $s$ in
Eq.~(\ref{alphadd})
is different from 2 due to the Frank-Condon factor for the relaxation
transition and
three-body dynamics.  

\begin{figure}
\label{relaxjila}
\centerline{\includegraphics[width=9cm]{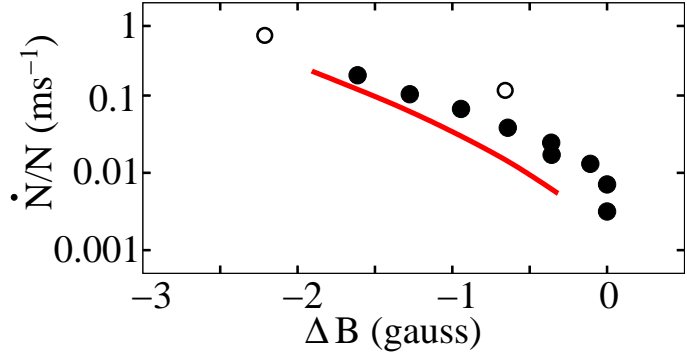}}
\caption{Two-body decay rate of a  $^{40}$K$_2$
ultracold molecular gas as a function of the magnetic field detuning from the
$^{40}$K Feshbach resonance at 202\,G. The dots indicate
experimental values, and the solid line shows the theoretical
results normalized to the experimental value at $\Delta B=-1.6\,$G.}
\end{figure}

\begin{figure}
\label{relax}
\centerline{\includegraphics[width=9cm]{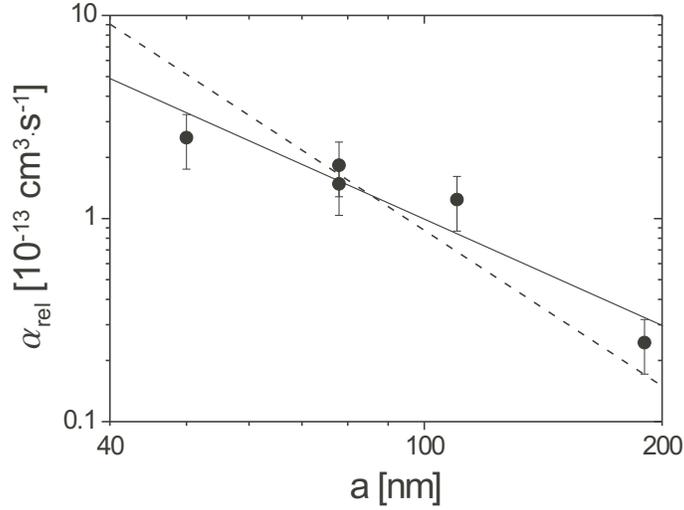}}
\caption{Two-body decay rate $\alpha_{rel}$ of a  $^6$Li$_2$
molecular condensate as a function of interatomic scattering length
near the $^6$Li Feshbach resonance at 834\,G. Solid line: least
square fit, $\alpha_{rel}\propto a^{-1.9\pm 0.8}$. Dashed line,
theory: $\alpha_{rel}\propto a^{-2.55}$. The theoretical
relaxation rate has been normalized to the experimental value at
$a=78\,$nm.}
\end{figure}

The remarkable collisional stability of weakly bound
molecules K$_2$ and Li$_2$ consisting of two fermionic atoms
has been observed in experiments at
JILA \cite{jila1,jila2,jila3}, Innsbruck \cite{rudy1,rudy2,rudy3},
MIT \cite{mit1,mit2}, ENS \cite{ens1,ens2}, Rice \cite{randy2}, and Duke \cite{john}. At
molecular densities $n\sim 10^{13}$ cm$^{-3}$ the lifetime of the
gas ranges from tens of milliseconds to tens of seconds, depending
on the value of the scattering length $a$. A strong decrease of
the relaxation rate with increasing $a$, following from
Eq.~(\ref{alphadd}), is  consistent with the experimental data. The
potassium experiment at JILA \cite{jila1} and lithium experiment
at ENS \cite{ens2} give the relaxation rate constant
$\alpha_{rel}\propto a^{-s}$, with $s\approx 2.3$  for K$_2$, and
$s\approx 1.9$ for Li$_2$, in agreement with theory ($s\approx
2.55$) within experimental uncertainty. The experimental and theoretical
results for potassium and lithium are shown in Fig.~5 and Fig~6. 
The absolute value of the rate
constant for a $^6$Li$_2$ condensate is $\alpha_{rel}\approx
1\times 10^{-13}$ cm$^3$/s for the scattering length $a\approx
110\,$nm. For K$_2$ it is an order of magnitude higher at the
same value of $a$ \cite{jila1}, which can be a consequence of a
larger value of the characteristic radius of interaction $R_e$.

The suppression of the relaxation decay rate of weakly bound
molecules of fermionic atoms has a crucial consequence for the
physics of these molecules. At realistic temperatures the
relaxation rate constant $\alpha_{rel}$ is much smaller than the
rate constant of elastic collisions $8\pi a_{dd}^2v_T$, where
$v_T$ is the thermal velocity. For example, for the Li$_2$ weakly
bound molecules at a temperature $T\sim 3\mu$K and $a\sim 800$\AA,
the corresponding ratio is of the order of $10^{-4}$ or $10^{-5}$.
This opens wide possibilities for reaching BEC of the molecules
and cooling the Bose-condensed gas to temperatures of the order of
its chemical potential. Long-lived BEC of weakly bound molecules
has been observed for $^{40}$K$_{2}$ at JILA
\cite{jila2,jila3} and for $^{6}$Li$_{2}$ at Innsbruck
\cite{rudy2,rudy3}, MIT \cite{mit1,mit2}, ENS \cite{ens2}, 
Rice \cite{randy2}, Duke \cite{john}, and recently at Melbourne \cite{hannaford}
and Tokyo \cite{ueda}. Measurements  of the molecule-molecule
scattering length confirm the result $a_{dd}=0.6a$ with the accuracy
up to 30\% \cite{rudy3,ens2}. 

\section{Heteronuclear molecules in Fermi-Fermi mixtures}

\subsection{Effect of mass ratio on elastic intermolecular
interaction}

We now focus on novel physics of heteronuclear molecules which are expected to be formed
in a mixture of two different fermionic atoms (Fermi-Fermi mixture) at a large
positive 2-body scattering length $a$.    
In several aspects, the physics is similar to that discussed above
for homonuclear molecules of fermionic atoms in different internal states.
However, for a large mass ratio of the atoms, the situation changes drastically. 
This is related to the existence of 3-body bound
Efimov states, which in general makes it impossible to describe
molecule-molecule scattering using only the value of the
2-body scattering length $a$. 

We start with calculating the amplitude of elastic interaction (scattering)
between
weakly bound heteronuclear molecules consisting of a heavy (mass $M$) and light
(mass $m$)
fermionic atoms, assuming that the atom-atom scattering length satisfies the
inequality
$a\gg R_e$ and again considering the ultracold limit determined by the
condition (\ref{ka}).
In this case the scattering is dominated by the contribution of the $s$-wave
channel, and we present
here the exact results obtained in Ref. \cite{PSSJ} using the zero-range
approximation. 
Under the condition $ka\ll 1$ the collision energy is much smaller than the molecular binding energy
$\varepsilon_0$. Hence, the $s$-wave molecule-molecule elastic scattering can
be
analyzed after setting the total energy equal  to $-2\varepsilon_0=-\hbar^2/\mu a^2$.
In the zero-range approximation one should solve the four-body free-particle
Schr\"odinger
equation which again can be written in the form (\ref{4bodySchr}):
\begin{eqnarray}  \nonumber
\left[-\nabla_{{\bf r}_1}^2-\nabla_{{\bf
r}_2}^2-\nabla_{\bf {R}}^2+2/a^2\right]\Psi =0,
\end{eqnarray}
where ${\bf r}_1$ is the distance between two given heavy and light fermions,
and
${\bf r}_2$ the distance between the other two (see Fig.~7). However, it is now
more
convenient to define the distance between the centers of mass of
these pairs as $\beta {\bf R}$, and the separations between the heavy and light
fermions in the other two possible heavy-light pairs as ${\bf r}_\pm=\alpha_\pm
{\bf
r}_1+\alpha_\mp{\bf r}_2\pm\beta {\bf R}$, with  
$\beta = \sqrt{2\alpha_+ \alpha_-}$, $\alpha_+=\mu/M$, $\alpha_-
=\mu/m$, and $\mu=mM/(m+M)$ being the reduced mass. 

\begin{figure}
\label{fig.coordinates}
\centerline{\includegraphics[width=7cm]{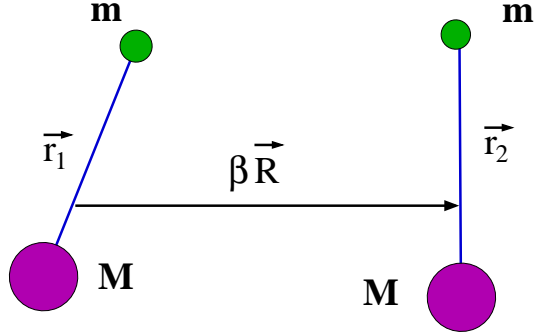}}
\caption{Set of coordinates for the four-body problem with two heteronuclear
molecules.}
\end{figure}

The symmetry condition (\ref{symmetry}) then takes the form:
\begin{eqnarray}\label{symmetryF}
\Psi({\bf r}_{1},{\bf r}_{2},{\bf {R}})=\Psi({\bf r}_{2},{\bf
r}_{1},-{\bf {R}})
=-\Psi ({\bf r}_\pm,{\bf r}_\mp,\pm\beta ({\bf r}_1-{\bf
r}_2)\mp(\alpha_+
-\alpha_-){\bf R}),
\end{eqnarray}
and the Bethe-Peierls boundary condition should be applied for a vanishing
distance in any pair of heavy and light fermions, i.e. for
${\bf r}_1\rightarrow 0$, ${\bf r}_2\rightarrow 0$, and ${\bf
r}_\pm\rightarrow 0$. For ${\bf r}_1\rightarrow 0$ it is again given 
by Eq.~(\ref{boundary}).

Due to the change in the definition of the coordinates the asymptotic expression for the 
wavefunction $\Psi$ at large distances $R$ now reads:
\begin{equation}   \label{asymptote1}
\Psi\approx\phi_0(r_1)\phi_0(r_2)(1-a_{dd}/\beta R);\,\,\,\,\,R\gg a,
\end{equation}
where the notation $a_{dd}$ is again used for the molecule-molecule scattering
length, 
and the wavefunction of a weakly bound molecule is given by Eq.~(\ref{phi0}). 
Then the asymptotic expression for the function $f({\bf r}_2,{\bf R})$ at large
$R$
is given by:
\begin{equation}\label{dimerdimer.swaveF}
f({\bf r}_2,{\bf R})\approx (2/r_2a)\exp{(-r_2/a)}(1-a_{dd}/\beta
R);\,\,\,\,R\gg a.
\end{equation}

For $s$-wave scattering the function $f$ depends only on
three variables: the absolute values of ${\bf r}_2$ and ${\bf {R}}$,
and the angle between them. Using the procedure described in Section~2
we obtain for $f$ the same integral equation (\ref{main}). 
The effect of different masses is contained in the expressions for the
vectors
$S_{\pm}$, which now read $S_\pm  =  \{\alpha_\mp {\bf r}' \pm \beta
{\bf
{R}}' ,\alpha_\pm {\bf r}' \mp\beta {\bf {R}}' , \mp\beta {\bf
r}' \mp(\alpha_+ -\alpha_-){\bf {R}}' \}$.

In order to find the molecule-molecule scattering length as a function of the
mass ratio $M/m$ it is again more convenient to transform the integral equation
for the function $f({\bf r},{\bf R})$ into an equation in the momentum space. Introducing
the Fourier transform of the function $f({\bf r},{\bf R})$ as $f({\bf k},{\bf p})=\int d^3
rd^3Rf({\bf r},{\bf R})\exp(i{\bf k\cdot r}/a+i\beta {\bf p\cdot R}/a)$, we obtain 
the following momentum-space equation:
\begin{eqnarray}\label{momentum}
\!\!\!\!\!&\!\!&\sum_\pm\!\!\int\!\!
\frac{f({\bf k}\pm \alpha_\mp ({\bf p}'-{\bf p}),{\bf p}')\,{\rm
d}^3p'}{2+\!\beta^2p'^2+\!({\bf k}\pm \alpha_\mp ({\bf p}'\!-{\bf
p}))^2+({\bf k}\pm
\alpha_\pm ({\bf p}'\!+{\bf p}))^2}\nonumber\\
\!\!\!\!&\!\!&\!=\!\int\!\! \frac{f({\bf k}',-{\bf p})\,{\rm
d}^3k'}{2\! +\! k'^2\!
+\! k^2\! +\! \beta^2p^2}- \frac{2\pi^2(1 +\! k^2\! +\beta^2p^2)f({\bf
k},{\bf
p})}{\sqrt{2 + k^2 +\beta^2p^2}+ 1}.
\end{eqnarray} 
By making the substitution $f({\bf k},{\bf p})=(\delta({\bf p})+g({\bf
k},{\bf
p})/p^2)/(1+k^2)$ we reduce Eq.~(\ref{momentum}) to an inhomogeneous
equation for
the function $g({\bf k},{\bf p})$. This equation is similar to Eq.~(\ref{g1}) and
we do not present it here because of its complexity. As well as in the case of $M=m$, 
for ${\bf p}\rightarrow 0$ the function $g({\bf k},{\bf p})$ tends to
a finite value
independent of ${\bf k}$. The molecule-molecule scattering length is again
given by $a_{dd}=-2\pi^2a\lim_{{\bf p}\rightarrow 0}g({\bf k},{\bf p})$. 
In Fig.~7 we display the ratio $a_{dd}/a$ versus the mass ratio $M/m$, found in Ref. \cite{PSSJ}.
For the case of homonuclear molecules ($m=M$) we recover the molecule-molecule
scattering length $a_{dd}=0.6a$. 

\begin{figure}
\label{fig.add}
\centerline{\includegraphics[width=10cm]{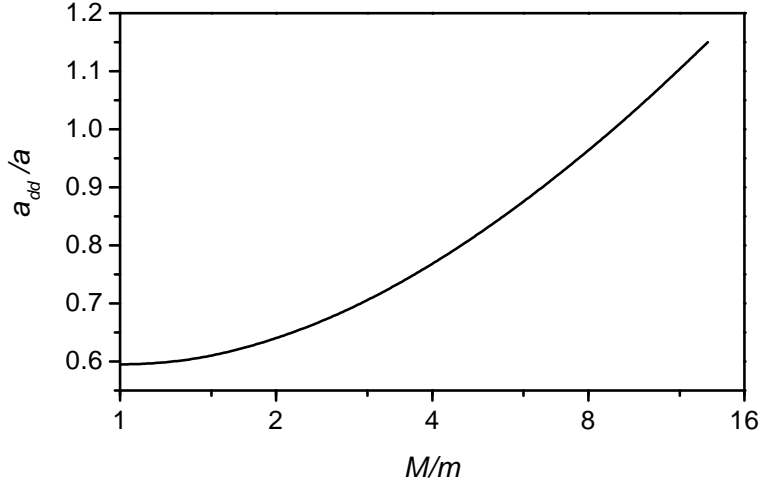}}
\caption{The ratio $a_{dd}/a$ versus $M/m$.}
\end{figure}

The universal dependence of $a_{dd}/a$ on the mass ratio, presented in
Fig.~8, can be established in the zero-range
approximation only if $M/m$ is smaller than 13.6. 
Calculations then show the absence of four-body weakly bound states,
and
for $M/m\sim 1$ the behavior of $f$ suggests a soft-core repulsion
between
molecules, with a range $\sim a$.
For the mass ratio larger than the limiting value 13.6, 
the description of the molecule-molecule scattering requires a
three-body parameter coming from the short-range behavior of the three-body
subsystem
consisting of one light and two heavy fermions \cite{efimov,Petrov2}.     
A qualitative explanation of this behavior will be given in subsection~\ref{BOP}.

\subsection{Collisional relaxation for moderate mass ratios}

The most exciting physics with weakly bound (heteronuclear) bosonic molecules 
consisting of different fermionic atoms is related to their collisional
stability.
As well as homonuclear molecules discussed above, they are in the
highest rovibrational state
and hence undergo relaxation into deeply bound states in
molecule-molecule collisions, which leads to decay of the sample. 
The collisional relaxation determines the lifetime of the Bose gas of
weakly bound molecules and there is a subtle question of whether and how
the mass ratio $M/m$ can influence the suppression of this process \cite{PSSJ},  
originating from the Fermi statistics for the atoms and playing a crucial role
in the case of homonuclear molecules.
In a similar way, behaving themselves as point-like bosons
at large intermolecular distances, heteronuclear molecules 
''remember'' that they consist of fermions when the intermolecular
separation becomes smaller than the molecule size ($\sim a$). 
The relaxation requires the presence of at least three fermions at 
separations $\sim R_e$
from each other. Two of them are necessarily identical, so that due to
the Pauli exclusion principle the relaxation probability acquires
a small factor proportional to a power of $(qR_e)$, where $q\sim
1/a$ is a characteristic momentum of the atoms in the weakly bound
molecular state. What changes in this picture when the fermionic atoms 
forming the molecule have different masses?

We first consider molecule-molecule relaxation collisions for the case 
where the mass ratio is smaller than the limiting value 13.6 and short-range 
physics is not supposed to influence the dependence of the relaxation 
rate on the 2-body scattering length $a$. As well as in the case of 
homonuclear molecules in Section~2, we assume the inequality $a\gg R_e$ 
and consider the ultracold limit described by Eq.~(\ref{ka}). 
The configuration space contributing to the relaxation
probability can be again viewed as a system of only three atoms at short
distances $\sim R_e$ from each other and a fourth atom separated from 
this system by a large distance $\sim a$. Hence, the four-body 
wavefunction decomposes into a product according to Eq.~(\ref{decomp}):
$\Psi=\eta({\bf z})\Psi^{(3)}(\rho,\Omega)$, with $\Psi^{(3)}$ being
the wavefunction of the three-fermion system, and $\rho\ll a$ and $\Omega$ 
being the hyperradius and the set of hyperangles for these fermions. 
The distance between their center of mass and the fourth atom is ${\bf z}$,
and the function $\eta({\bf z})$ describes the motion of this atom. 
In the case of fermionic atoms with different masses there are two
possible choices
of a three-body subsystem out of four fermions. The most important is
the relaxation 
in the system of one atom with the mass $m$ and two heavier atoms with
masses $M$.  

We then use the same arguments as in Section~2 and obtain 
equation (\ref{Psi3nu}) for the function $\Psi^{(3)}$ at distances where
$R_e\ll \rho\ll a$:
$\Psi^{(3)}=A(a)\Phi_{\nu}(\Omega)\rho^{\nu-1}$,
with the coefficient $A(a)$ determining the $a$-dependence of the relaxation
rate 
according to Eq.~(\ref{alphaA}): $\alpha_{rel}=\alpha^{(3)}\propto |A(a)|^2$.
A similar scaling procedure as in Section~2 again leads to 
$\alpha_{rel}\propto a^{-s}$, where $s=2\nu+1$, and restoring the dimensions
we can write the relaxation rate in the form (\ref{alphadd}):
$\alpha_{rel}= C(\hbar R_e/m)(R_e/a)^s$, 
with a coefficient $C$ depending on the mass ratio and on short-range physics. 

However, the exponent $s$ now depends not only on the symmetry of the
three-body wave function $\Psi^{(3)}$, but also on the mass ratio $M/m$. 
The smallest value of $\nu$, i.e. the one corresponding to the leading 
relaxation channel at large $a$, is achieved for the $p$-wave symmetry 
in the system of one light and two heavy fermions \cite{Petrov2}. 
In the interval $-1\leq\nu<2$ it is
given by the root of the function \cite{Petrov2}:
\begin{equation}\label{lambda}
\lambda(\nu)=\frac{\nu(\nu+2)}{\nu+1}\cot\frac{\pi\nu}{2}+\frac{\nu\sin\gamma\cos(\nu\gamma+
\gamma)-\sin(\nu\gamma)}{(\nu+1)\sin^2\gamma\cos\gamma\sin(\pi\nu/2)}
\end{equation}
where $\gamma=\arcsin\left[M/(M+m)\right]$. A detailed derivation of 
Eq.~(\ref{lambda}) is given in Ref. \cite{Petrov2}.

In the case of equal masses we recover $s=2\nu+1\approx 2.55$ obtained in
Section~2, and it slowly decreases with increasing the
mass ratio (see Fig.~9). For $M/m\sim 1$ nothing dramatic happens: the
suppression of the relaxation 
rate with increasing the 2-body scattering length $a$ becomes slightly weaker 
than for homonuclear molecules. However, for the mass ratio approaching the
limiting value 13.6 the exponent $s$ first reaches zero and then becomes
negative,
showing even an increase of the relaxation rate with $a$. We will give a
qualitative
explanation of this phenomenon using the Born-Oppenheimer approximation 
for the system of two heavy and one light atom.

\begin{figure}
\label{fig.nu}
\centerline{\includegraphics[width=10cm]{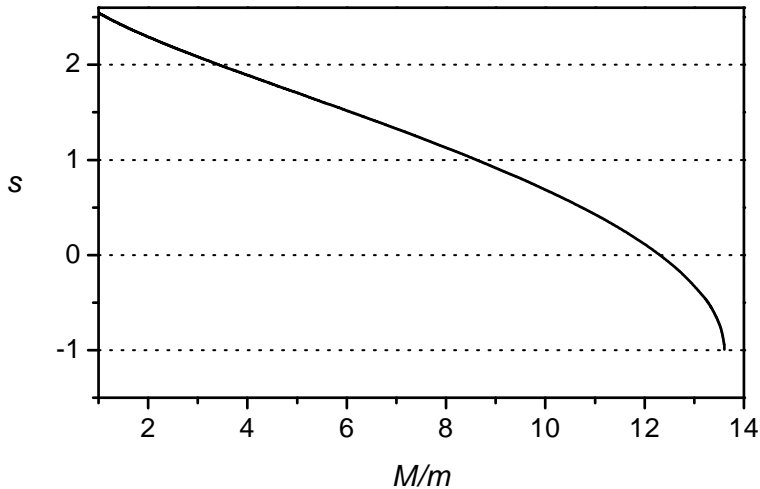}}
\caption{The dependence of the exponent $s=2\nu+1$ in
Eq.~(\ref{alphadd}) on the mass
ratio $M/m$.}
\end{figure}

\subsection{Born-Oppenheimer picture of collisional relaxation}
\label{BOP}

In the Born-Oppenheimer approximation one assumes that the state of a fast light atom
adiabatically adjusts itself to the positions ${\bf R}_1$ and ${\bf R}_2$ of the 
slow heavy atoms. One then finds the wavefunction and energy of a bound state 
of the light atom with two heavy atoms at a given separation between them, 
${\bf R}={\bf R}_1-{\bf R}_2$. For convenience, from this point on we change the notations and use
${\bf R}$ to specify the distance between the heavy atoms, and ${\bf r}$ the coordinate of a light atom relative to their
center of mass. In general, there are two states of a light atom in the field of two heavy ones: the gerade state 
$(+)$ with the wavefunction remaining unchanged under permutation of the heavy atoms $({\bf R}\rightarrow -{\bf R})$,
and the ungerade state $(-)$ with the wavefunction changing its sign under this operation. The corresponding wavefunctions are given by
\begin{equation}       \label{psi3}
\psi_{\bf R}^{\pm}({\bf r}) = {\cal N}_\pm \left( \frac{e^{- \kappa_{\pm}(R)|{\bf r}-{\bf R}/2|}}{|{\bf r}-{\bf R}/2|} 
\pm \frac{e^{-\kappa_{\pm}(R)|{\bf r}+{\bf R}/2|}}{|{\bf r}+{\bf R}/2|} \right),
\end{equation}
where ${\cal N}_\pm$ are normalization coefficients which depend on $R$. 
The corresponding binding energies are  
\begin{equation}\label{BOBindingEnergies}
\epsilon_{\pm} (R) = -\hbar^2\kappa_{\pm}^2(R)/2 m,
\end{equation}
and the parameters $\kappa_{\pm}(R)$ follow from the equation
\begin{equation} \label{kappapm}
\kappa_{\pm}(R) \mp \exp \left[-\kappa_{\pm}(R)R \right]/R = 1/a.
\end{equation}
Equation (\ref{kappapm}) is obtained by using the Bethe-Peierls 
boundary condition (\ref{boundary1}) for the wavefunctions $\psi^{\pm}_{\bf R}$ of Eq.~(\ref{psi3}) at vanishing light-heavy atom separations
$|{\bf r}\pm{\bf R}/2|$.

The ungerade $(-)$ state energy is always higher than the energy of the gerade $(+)$ 
state. Moreover, for $R<a$ the ungerade state is no longer bound and we are dealing only with the gerade bound state.
In the limit of $R\ll a$ Eq.~(\ref{kappapm}) gives $\kappa_+=0.56$. Then the energy of the gerade bound state,
representing an effective potential for the relative motion of the heavy atoms, is given by 
\begin{equation}   \label{epsilonpluslimit}
\epsilon_+(R)=-0.16\hbar^2/mR^2.  
\end{equation}

We thus see that when the heavy atoms are separated from each
other by a distance $R\ll a$, the light atom mediates an effective $1/R^2$
attraction between them.
Actually, the same result follows from the Efimov picture of effective interaction in a
three-body system \cite{efimov} and the Born-Oppenheimer approximation only
gives a physically transparent illustration of this picture
\cite{PSSJ,Fonseca}.
For a large mass ratio the mediated attractive potential 
$\epsilon_+(R)=-0.16\hbar^2/mR^2$ strongly modifies the physics of the relaxation 
process. It competes with the Pauli principle which in terms of
effective interaction manifests itself in the centrifugal $1/R^2$ repulsion 
between the heavy atoms.  
The presence of this repulsion is clearly seen from the fact that the
light-atom wavefunction $\psi^+_{{\bf R}}({\bf r})$ does not change sign under 
permutation of heavy fermions. As the total wavefunction of the three-body 
system $\psi^+_{{\bf R}}({\bf r})\chi({\bf R})$ is antisymmetric with respect to 
this permutation, the wavefunction of the relative
motion of heavy atoms $\chi({\bf R})$ should change its sign. Therefore,
$\chi({\bf R})$ contains only 
partial waves with odd angular momenta, and for the lowest angular
momentum ($p$-wave) 
the centrifugal barrier is $U_c(R)=2\hbar^2/MR^2$. For comparable
masses it is significantly 
stronger than $\epsilon_+(R)$. Thus, we have the physical picture discussed in the
case of homonuclear molecules: the Pauli
principle (centrifugal barrier) reduces the probability for the
atoms to be at short distances and, as a consequence, the relaxation
rate decreases
with increasing the atom-atom scattering length $a$.

The role of the effective attraction increases with $M/m$. As a
result, the decrease of the relaxation rate with increasing $a$ becomes 
weaker.  The exponent $s$ in Eq.~(\ref{alphadd}) continuously decreases with
increasing $M/m$
and becomes zero for $M/m=12.33$ (see Fig.~9). In the Born-Oppenheimer picture
this means that
at this point one has a balance between the mediated attraction and the
centrifugal
repulsion. A further increase in $M/m$ makes $s$ negative and it reaches the
value $s=-1$ for the critical mass ratio $M/m=13.6$. Thus, in the range
$12.33<M/m<13.6$ the relaxation rate increases with $a$.

For an overcritical mass ratio $M/m>13.6$ we have a well-known
phenomenon of the fall
of a particle to the center in an attractive $1/R^2$ potential
\cite{LL3}. In this case the shape of
the wavefunction at distances of the order of $R_e$ can significantly influence
the
large-scale behavior and a short-range three-body parameter is
required to describe the system. The wavefunction of heavy atoms $\chi({\bf
R})$ 
acquires many nodes at short distances $R$, which indicates the appearance of
3-body bound Efimov states.

\subsection{Molecules of heavy and light fermionic atoms}

The discussion of the previous subsection shows that weakly 
bound molecules of heavy and light fermions become collisionally unstable for
the mass
ratio $M/m$ close to the limiting value 13.6. The effect of the Pauli principle 
becomes weaker than the attraction between heavy atoms at distances $R\ll a$,
mediated by light fermions. However, this picture explains only the dependence
of the relaxation rate on the 2-body scattering length $a$. At the same time, for 
heteronuclear molecules the relaxation rate and the amplitude of elastic 
molecule-molecule interaction can also depend on the mass ratio 
irrespective of the value of $a$ and short-range physics. To elucidate  
this dependence, we will look at the
interaction between the molecules of heavy and light fermions at large 
intermolecular separations.   

We consider the interaction between two such molecules in the
Born-Oppenheimer approximation and calculate the wavefunctions and binding
energies of two light fermions in the field of two heavy atoms fixed at their positions
${\bf R}_1$ and ${\bf R}_2$. The sum of the
corresponding binding energies gives an effective interaction
potential $U_{\rm eff}$ for the heavy fermions as a function of the separation 
$R=|{\bf R}_1-{\bf R}_2|$ between them. 

For $R>a$, there are two bound states, gerade (+) and ungerade (-), for a light atom interacting with
a pair of fixed heavy atoms. Their wavefunctions are given by Eq.~(\ref{psi3}), and the corresponding 
binding energies follow from Eqs.~(\ref{BOBindingEnergies}) and (\ref{kappapm}). For large $R$ satisfying the
condition $\exp(-R/a)\ll 1$, Eq.~(\ref{BOBindingEnergies}) yields:
\begin{equation}\label{BOBELargeR}
\epsilon_{\pm} (R) \approx -|\epsilon_0|\mp 2|\epsilon_0|\frac{a}{R}\exp(-R/a)+\frac{U_{\rm ex}(R)}{2},
\end{equation}
where the binding energy of a single molecule, $\epsilon_0$, is given by Eq.~(\ref{binding}) with the reduced mass
$\mu$ very close to the light atom mass $m$.  
\begin{equation}\label{Exchange}
U_{\rm ex}(R)=4|\epsilon_0|\frac{a}{R}\left(1-\frac{a}{2R}\right)\exp(-2R/a).
\end{equation}

Since the light fermions are identical, their two-body wavefunction is
an antisymmetrized product of gerade and ungerade wavefunctions:
\begin{equation}\label{Antisymm}
\psi_{\bf R}({\bf r_1,r_2})=[\psi_{\bf R}^{+}({\bf r_1})\psi_{\bf R}^{-}({\bf r_2})-\psi_{\bf R}^{+}({\bf r_2})\psi_{\bf R}^{-}({\bf r_1})]/\sqrt{2}.
\end{equation}
The Born-Oppenheimer adiabatic approach is valid at distances $R>a$, where the effective interaction potential between the molecules, $U_{\rm eff}$, is the sum of 
$\epsilon_{+}(R)$ and $\epsilon_{-}(R)$ (one should add $2|\epsilon_0|$ so that $U_{\rm eff}(R)\rightarrow 0$ for $R\rightarrow\infty$). This potential is displayed in Fig~10.
\begin{figure}
\label{fig15}
\centerline{\includegraphics[width=10cm]{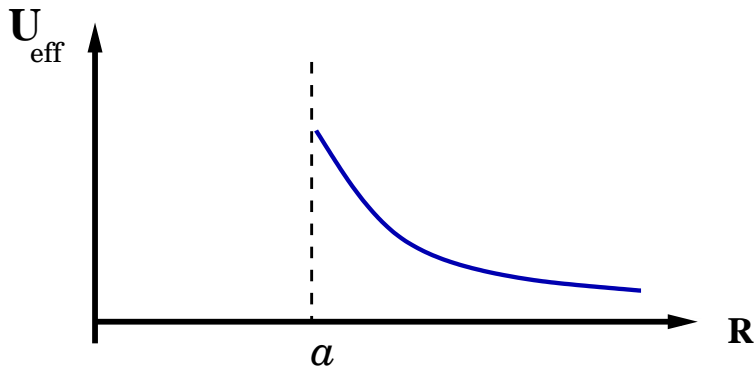}}
\caption{Interaction potential for two molecules of heavy and light fermions as
a function
of the separation $R$ between the heavy atoms.}
\end{figure}

At sufficiently large inter-heavy separations where Eq.~(\ref{BOBELargeR}) is valid, the effective
potential can be written as 
\begin{equation}\label{BOUeffective}
U_{\rm eff}(R)=\epsilon_{+}(R)+\epsilon_{-}(R)+2|\epsilon_0|\approx U_{\rm ex}(R).
\end{equation} 
The potential $U_{\rm ex}$ originates from the exchange of light fermions and thus can be treated as an exchange interaction. It is purely 
repulsive and, according to Eq.~(\ref{Exchange}), has the asymptotic shape of a Yukawa potential at large $R$. Direct calculations show that $U_{\rm ex}$ 
is a very good approximation to $U_{\rm eff}$ for $R \gtrsim 1.5 a$.

We now demonstrate the calculation of the dimer-dimer scattering length $a_{dd}$ in the limit of $M/m\gg 1$ \cite{marcelis}. 
In the Born-Oppenheimer approach the Schr\"odinger equation for the relative motion of two molecules reads;
\begin{equation}\label{BOSchr}
( -(\hbar^2/m)\nabla^2_{\bf R}+U_{\rm eff}(R) -\epsilon)\Psi({\bf R})=0,
\end{equation}
where $\epsilon$ is the collision energy. Note, that the repulsive effective potential is inversely proportional to 
the light mass $m$, whereas the kinetic energy operator in Eq.~(\ref{BOSchr}) has a prefactor $1/M$. Therefore, for a large mass ratio 
$M/m$, the heavy atoms approach each other at distances smaller than $a$ with an exponentially small tunneling probability $P\propto \exp (-B\sqrt{M/m})$, 
where $B\sim 1$. This leads to the relaxation rate constant
\begin{equation}   \label{alphaMm}
\alpha_{rel}\propto\exp (-B\sqrt{M/m}),
\end{equation}
which strongly decreases with increasing the mass ratio $M/m$. 

The analysis shows that the elastic part of the scattering amplitude can be calculated with a very high accuracy using Eq.~({\ref{BOSchr}) 
for $M/m\gtrsim 20$ and is practically insensitive to the way we choose the boundary condition for the wavefunction at $R=a$. 
The dominant contribution to the scattering 
comes from distances in the vicinity of $R=a_{dd}\gg a$, where the effective potential can be approximated by Eq.~(\ref{Exchange}) with a constant preexponential factor:
\begin{equation}\label{BOExp}
U_{\rm eff}(R)\approx 2\hbar^2(ma a_{dd})^{-1}\exp(-2R/a).    
\end{equation}
Then, the zero energy solution of Eq.~(\ref{BOSchr}) that decays at smaller $R$, reads:
\begin{equation}\label{BOApproxSol}
\Psi(R) = \frac{a}{R}K_0\left(\sqrt{\frac{2M}{m}\frac{a}{a_{dd}}}e^{-R/a}\right),    
\end{equation}
where $K_0$ is the decaying Bessel function. Comparing the result of Eq.~(\ref{BOApproxSol}) at large $R$ with the asymptotic behavior $\Psi(R) \propto (1-a_{dd}/R)$ 
we obtain an equation for $a_{dd}$:
\begin{equation}\label{addM}
a_{dd}=\frac{a}{2}\ln \left(\frac{e^{2\gamma}}{2} \frac{M}{m}\frac{a}{a_{dd}}\right).
\end{equation}
This gives 
\begin{equation}   \label{addsimple}
a_{dd}\approx a\ln \sqrt{M/m}, 
\end{equation}
and the scattering cross section is
\begin{equation}    \label{sigmadd}
\sigma_{dd}=8\pi a_{dd}^2.
\end{equation}
From Eq.~(\ref{BOApproxSol}) we see that the interval of distances near $R=a_{dd}$, where the wavefunction changes, is of the order of $a$. This justifies 
the use of Eq.~(\ref{BOExp}). In fact, the corrections to Eq.~(\ref{addM}) can be obtained by treating the 
difference between Eqs.~(\ref{Exchange}) and (\ref{BOExp}) perturbatively. In this way the first order correction to the dimer-dimer scattering length is 
$-(3/4)a^2/a_{dd}$, where $a_{dd}$ is determined from Eq.~(\ref{addM}).

Qualitatively, $U_{\rm eff}(R)$ can be viewed as a hard-core potential with the radius $a_{dd}$, where the edge is smeared out on a lengthscale 
$\sim a\ll a_{dd}$. Therefore, the ultracold limit for dimer-dimer collisions, required for the validity of Eq.~(\ref{sigmadd}), is realized for relative
momenta of the dimers, $k$, satisfying the inequality
\begin{equation}   \label{ul}
ka_{dd}\ll 1.
\end{equation}
It can be useful (see Ref.~\cite{PAPSS}) to approximate the potential $U_{\rm eff}$ by a pure 
hard core with the radius $a_{dd}$. This approximation works under the condition $ka\ll 1$, which is less strict than Eq.~(\ref{ul}).

Let us now mention that numerical calculations Ref. \cite{marcelis} show that there are no resonances in the dimer-dimer scattering amplitude, which
could appear in the presence of a weakly bound state of two dimers. Here we give a qualitative explanation of the absence of these bound states. Suppose 
there is such a state with energy $\epsilon\rightarrow 0$. Then, at distances $R>a$ the wavefunction of the heavy atoms should exponentially decay on the 
distance scale $\sim a\sqrt{m/M}\ll a$, since $U_{\rm eff}$ represents a barrier with the height $\sim 1/ma^2$. This means that the heavy atoms in such 
a bound state should be localized mostly at distances smaller than $a$. The gerade light atom is also localized at these distances as seen from the shape 
of the function $\psi^{+}$. The motion of the ungerade light atom relative to the localized trimer can be viewed as scattering with odd values of 
the angular momentum, and due to the centrifugal barrier the bound states of this atom with the trimer should be localized at distances $\sim a$
from the heavy atoms. In this case one would expect the BO approximation to work, since the ungerade light atom is moving much faster than the heavy 
atoms. However, this leads to a contradiction, because in the BO approach discussed above the ungerade state at interheavy separations $R<a$ is unbound.
We thus conclude that weakly bound states of two dimers are absent.

Although there are no resonances in the dimer-dimer collisions, there are branch-cut singularities in the scattering amplitude. They are related to the presence of inelastic
processes in molecule-molecule collisions. These represent the relaxation of one of the colliding dimers into a deeply bound state, 
the other dimer being dissociated, and the formation of bound trimers consisting of two heavy and one light atom, the other light atom carrying away 
the released binding energy.

\subsection{Trimer states}

The trimer states, which in most cases can be called Efimov trimers, are interesting objects. 
Their existence can be seen from the BO picture for two heavy atoms and one light atom in the gerade state. 
Within the BO approach the three-body problem reduces to the calculation 
of the relative motion of the heavy atoms in the effective potential created by the light atom. 
For the light atom in the gerade state, this potential is $\epsilon_+(R)$ 
found in the previous subsection. The Schr\"odinger equation for the wavefunction of the relative motion 
of the heavy atoms, $\chi_{\nu}({\bf R})$, reads:
\begin{equation}\label{BO3body}
\hat H\chi({\bf R})=\left[ -(\hbar^2/M)\nabla^2_{\bf R}+\epsilon_{+}(R)\right]\chi_\nu({\bf R}) =\epsilon_\nu\chi_\nu({\bf R}).
\end{equation}
The trimer states are nothing else than the bound states of heavy atoms in the effective potential $\epsilon_{+}(R)$. Accordingly, 
they correspond to the discrete part of the spectrum $\epsilon_{\nu}$, where the symbol $\nu$ denotes a set containing 
angular ($l$) and radial ($n$) quantum numbers. For $R\ll a$ the potential $\epsilon_{+}(R)$ 
is proportional to $-1/R^2$ (see Eq.~(\ref{epsilonpluslimit})) and, if this effective attraction overcomes the centrifugal 
barrier, we arrive at the well known phenomenon of the 
fall of a particle to the center in an attractive $1/R^2$ potential. 
Then, for a given orbital angular momentum $l$, the radial part of $\chi_\nu$ can be written as
\begin{equation} \label{TrimerSmallR}
\chi_\nu(R) \propto R^{-1/2}\sin(s_l \ln{R/r_0}), \quad R \ll a,
\end{equation} 
where 
\begin{equation}\label{sl}
s_l=\sqrt{0.16M/m-(l+1/2)^2}.
\end{equation} 
The three-body parameter $r_0$ determines the phase of the wavefunction at small distances and, in principle, depends on $l$. The wavefunction (\ref{TrimerSmallR}) has infinitely
many 
nodes, which means that in the zero-range approximation there are infinitely many trimer states. This is one of the properties of three-body systems with resonant interactions 
discovered by Efimov \cite{efimov}. We see that the fall to the center is possible in many angular momentum channels, provided the mass ratio is sufficiently large. However, 
for practical purposes and for simplicity, it is sufficient to consider the case where the Efimov effect occurs only for the angular momentum channel with the lowest possible 
$l$ for a given symmetry. This implies that when the heavy atoms are fermions and one has odd $l$, in order to confine ourselves to $l=1$ we should have the mass ratio in the 
range $14\lesssim M/m \lesssim 76$. For bosonic heavy atoms where $l$ is even, we set $l=0$ and consider $M/m\lesssim 39$ to avoid the Efimov effect for $l\geq 2$. 
In both cases we need a single three-body parameter $r_0$.

The formation of Efimov trimers in ultracold dimer-dimer collisions is energetically allowed only if $\epsilon_\nu<-2|\epsilon_0|$. 
This means that the trimers that we are interested in are relatively well bound and their size is smaller than $a$. Therefore, the process of the trimer formation is exponentially 
reduced for large mass ratios as the heavy atoms have to tunnel under the repulsive barrier $U_{\rm eff}$(R). Moreover, this process requires all of the four atoms to approach 
each other at distances smaller than $a$, and its rate decreases with the trimer size because it is more difficult for two identical light fermions to be in a smaller 
volume.

From Eq.~(\ref{TrimerSmallR}) one sees that the behavior of the three-body system does not change if $r_0$ is multiplied by
\begin{equation}\label{lambdal}
\lambda_l=\exp(\pi /s_l).
\end{equation} 
On the other hand, the dimensional analysis shows that the quantity $\epsilon_\nu/\epsilon_0$ depends only on the ratio $a/r_0$. This means that except for a straightforward
scaling with $a$, properties of the three-body system do not change when $a$ is multiplied or divided by $\lambda_l$. This discrete scaling symmetry of a three-body system, 
which shows itself in the log-periodic dependence of three-body observables, has yet to be observed experimentally. In the case of three identical bosons, where the BO approach
does not work and it is necessary to solve the three-body problem exactly \cite{efimov}, the observation of the consequences of the discrete scaling requires to change $a$ by a
factor of 
$\lambda\approx 22.7$, which is technically very difficult in ongoing experiments with cold atoms. In this respect three-body systems with a very large mass difference 
can be more favorable because of smaller values of $\lambda$. For example, in order to see one period of the log-periodic dependence in a Cs-Cs-Li three-body system $a$ has 
to be changed only by a factor of $\lambda \approx 5$.

At this point it is worth emphasizing that three-body effects can be observed in a gas of light-heavy dimers, where the interdimer repulsion originating from the exchange 
of the light fermions strongly reduces the decay rate associated with relaxation of the dimers into deep bound states. 
The trimer formation in dimer-dimer collisions is very sensitive to the positions and sizes of Efimov states, and the measurement of the formation rate can be used to 
demonstrate the discrete scaling symmetry of a three-body system. Indeed, this rate should have the log-periodic dependence on $a$ and is detectable by measuring the
lifetime of the gas of dimers. 

Besides the Efimov trimers, one can have  of one light and two heavy atoms may form ``universal'' trimer states well described in the zero-range approximation without introducing
the three-body parameter \cite{Kartavtsev}. In particular, they exist for the orbital angular momentum $l=1$ and mass ratios below the critical value, where the Efimov 
effect is absent and short-range physics drops out of consideration. One of such states emerges at $M/m \approx 8$ and crosses the trimer formation threshold 
($\epsilon_{tr}=-2|\epsilon_0|$) at $M/m\approx 12.7$. The existence of this state is already seen in the BO picture. It appears 
as a bound state of fermionic heavy atoms in the potential $\epsilon_+(R)$ for $l=1$. The other state exists at $M/m$ even closer to the critical mass ratio and never becomes
sufficiently deeply bound to be formed in cold dimer-dimer collisions. The universal
trimer states also exist for $l>1$ and $M/m>13.6$ \cite{Kartavtsev}. However, the trimer formation in dimer-dimer collisions at such mass ratios is dominated by the contribution
of Efimov trimers with smaller $l$. Therefore, below we focus on the formation of Efimov trimers. 

The calculation of the intrinsic lifetime of a trimer requires a detailed knowledge of short-range physics and is a tedious task.
Estimates of the imaginary part of the trimer energy, $\tau^{-1}$, from the experimental data on Cs$_3$ trimers \cite{kraemer06} show that it is approximately by a factor of 4
smaller than the real part $\epsilon_{\nu}$ (in this case $\eta_*\approx 0.06$). From a general point of view, we do not expect that the trimers with a binding energy
$\epsilon_{\nu}<-2|\epsilon_0|$ are very long-lived. However, one can have relatively narrow resonances, and we demonstrate calculations for various values of the elasticity
parameter $\eta_*$ \cite{marcelis}.

\subsection{Collisional relaxation of molecules of heavy and light fermions and formation of trimers} 

Let us now discuss inelastic processes in dimer-dimer collisions and start with relaxation of the dimers into deeply bound states. The typical size of a deeply bound state is of
the
order of the characteristic radius of the corresponding interatomic potential. We first consider the relaxation channel that requires one light and two heavy atoms to approach each
other at distances $\sim R_e\ll a$. 
Unlike the trimer formation, this decay mechanism is a purely three-body process. The other light atom is just a spectator. A qualitative scenario of this process is the following. 
With the tunneling probability which is exponentially suppressed for large $M/m$, two dimers approach each other at distances $R\sim a$. Then the heavy atoms are accelerated 
towards each other in the potential $\epsilon_{+}(R)$, and the light atom in the gerade state is always closely bound to the heavy ones as is seen from the shape of the function 
$\psi^{+}$. The relaxation transition occurs when the heavy atoms (and the gerade light fermion) are at interatomic separations 
$\sim R_e$. The relaxation rate constant certainly satisfies Eq.~(\ref{alphaMm}), but we should also find out how to take into account 
the relaxation process in the description of the trimer states and their  formation. The most convenient way to do so is to consider the three-body parameter $r_0$ as a complex
quantity and introduce the so-called elasticity parameter 
$\eta_*=-s_l \mathrm{Arg} (r_0)$ \cite{braaten07}. As follows from the asymptotic expression for the wavefunction (\ref{TrimerSmallR}), 
a negative argument of $r_0$ ensures that the incoming flux of heavy atoms is not smaller than the outgoing one:
\begin{equation}\label{Imr0}
\Phi_{\rm out}/\Phi_{\rm in}=\exp[4 s_l \mathrm{Arg} (r_0)] = \exp[-4\eta_*]\leq 1.
\end{equation}
This mimics the loss of atoms at small distances due to the relaxation into deeply bound states. In the analysis of Efimov states, the imaginary part of $r_0$ leads to 
the appearance of an imaginary part of $\epsilon_\nu$. This means that any Efimov state has a finite lifetime $\tau$ due to the relaxation. 
For small $|\mathrm{Arg} (r_0)|$ and for trimer states that are localized at distances smaller than $a$, we get $\tau^{-1}/|\epsilon_{\nu}|=4|\mathrm{Arg} (r_0)|=4\eta_*/s_l$. 
Strictly speaking this fact indicates that it is not possible to separate the relaxation process from the trimer formation  
because the trimers that are formed in dimer-dimer collisions will eventually decay due to the relaxation. Nevertheless,  
both the modulus and the argument of the three-body parameter can be determined by measuring the lifetime of a gas of dimers, leading to a number of quantitative 
predictions concerning the structure of Efimov states in the three-body subsystem of one light and two heavy atoms.

Another relaxation channel is the one in which two light atoms approach a heavy atom at distances $\sim R_e\ll a$. This channel is, however, suppressed
due to the Fermi statistics for the light atoms, which strongly reduces the probability of having them in a small volume. As a result, for realistic parameters
this relaxation mechanism is much weaker than the one in the system of one light and two heavy atoms \cite{marcelis}.

The study of the formation of trimer states in molecule-molecule collisions requires us to go beyond the conventional Born-Oppenheimer approximation,
since this approximation breaks down for the ungerade light atom at separations between the heavy atoms $R<a$. Within the recently developed ``hybrid Born-Oppenheimer'' approach
\cite{marcelis} 
one applies the Born-Oppenheimer method to the gerade light fermion which is characterized by the wavefunction $\psi^{+}_{\bf R}({\bf r})$ and energy $\epsilon_+(R)$ adiabatically
ajusting themselves to the motion of the heavy atoms. The gerade light atom is then integrated out by introducing the potential $\epsilon_+(R)$ for the heavy atoms. Once
this is done the original 4-body problem is reduced to a 3-body problem described by the Schr\"odinger equation equation  
\begin{equation}\label{Reduced3body}
[\hat{H}-(\hbar^2/2\mu_3)\nabla_{\bf r}^2-E]\Psi({\bf R},{\bf r})=0,
\end{equation}
where $\hat{H}$ is given by Eq.~(\ref{BO3body}), $\mu_3=2mM/(2M+m)$, $E=-2|\epsilon_0|+\epsilon$ is the total energy of the four-body system in the center of mass reference frame, 
and $\epsilon$ is the dimer-dimer collision energy. This problem is then treated exactly in the Bethe-Peierls approach, and the interaction of the light atom with the heavy atoms
is included in the form of the Bethe-Peierls boundary condition (\ref{boundary1}) for $\Psi$ at vanishing light-heavy separations $|{\bf r}\pm {\bf R}/2|$. The ungerade symmetry
for this atom is taken into account by the condition 
\begin{equation}\label{CondUngerade}
\Psi({\bf R},{\bf r})=-\Psi({\bf R},-{\bf r}).
\end{equation}
As the heavy atoms are identical fermions, 
we have $\Psi({\bf R},{\bf r})=-\Psi(-{\bf R},{\bf r})$. Combined with Eq.~(\ref{CondUngerade}), this leads to the condition 
$\Psi({\bf R},{\bf r})=\Psi(-{\bf R},-{\bf r})$. Therefore, $\Psi({\bf R},{\bf r})$ describes atom-dimer scattering with even angular momenta, and for 
ultracold collisions we have to solve an $s$-wave atom-dimer scattering problem. 

In order to solve Eq.~(\ref{Reduced3body}) we follow the method of Ref. \cite{Petrov2} and introduce an auxiliary function $f({\bf R})$ and write down the 
wavefunction $\Psi({\bf R},{\bf r})$ in the form:
\begin{equation} \label{fourbodysolution}
\Psi({\bf R,r})=\sum_{\nu}\int_{\bf R'} \chi_{\nu}({\bf R}) \chi_{\nu}^{\ast}({\bf R'})K_{\kappa_\nu}(2 {\bf r},{\bf R'})f({\bf R'}),
\end{equation}
where
\begin{equation}\label{Knu}
K_{\kappa_\nu}(2 {\bf r},{\bf R'})=\frac{e^{-\kappa_\nu |{\bf r}-{\bf R'}/2|}}{4\pi |{\bf r}-{\bf R'}/2|}-\frac{e^{-\kappa_\nu |{\bf r}+{\bf R'}/2|}}{4\pi |{\bf r}+{\bf R'}/2|}
\end{equation}
and
\begin{equation}\label{kappa}
\kappa_\nu=\left\{ \begin{array}{ll} \sqrt{2\mu_3(\epsilon_\nu-E)}/\hbar,& \epsilon_\nu>E,\\
-i\sqrt{2\mu_3(E-\epsilon_\nu)}/\hbar,& \epsilon_\nu<E. \end{array} \right.
\end{equation}

For $\epsilon_\nu<E$, the trimer can be formed in the state $\nu$. In this case $\kappa_\nu$ is imaginary and the function (\ref{Knu}) describes an outgoing wave of 
the light atom moving away from the trimer. The choice of the sign in Eq.~(\ref{kappa}) ensures that there is no incoming flux in the atom-trimer channel.  

Using the Bethe-Peierls boundary condition (\ref{boundary1}) for the wavefunction (\ref{fourbodysolution}) at $|{\bf r}\pm {\bf R}/2|\rightarrow 0$ we one obtain an integral
equation for the function $f({\bf R})$:
\begin{equation}\label{final}
\left\{\hat{L}-\hat{L}'+\sin^2\theta\frac{\sqrt{2\mu(\epsilon_0-E)}/\hbar -1/a}{4\pi}\right\}f({\bf R})=0,
\end{equation}
where $\mu=mM/(m+M)$, $\theta=\arctan{\sqrt{1+2M/m}}$, and 
\begin{equation}\label{Lprime}
\!\!\!\!\!\!\hat{L}'f({\bf R})\!\!=\!\!\!\int_{\bf R'} \!\sum_{\nu} [\chi_{\nu}({\bf R}) \chi_{\nu}^{\ast}({\bf R'})K_{\kappa_\nu}({\bf
R,R'})\!-\!\chi_{\nu}^0({\bf R}) 
\chi_{\nu}^{0\ast}({\bf R'})K_{\kappa_\nu^0}({\bf R,R'})]f({\bf R'}),
\end{equation}
\begin{equation}\label{L}
\!\!\!\!\!\!\!\hat{L} f({\bf R})\!\!=\!\!P\!\!\!\int_{\bf R'}\!\! \left\{\!\!G(|{\bf R\!-\!R'}|)\left[ f({\bf R})\!\!-\!\!f({\bf R'}) \right]\!\pm\! G(\sqrt{R^2\!+\!R'^2\!-
\!\!2{\bf
R\!\!\cdot
\!\!R'}\cos{2\theta}})f({\bf R'})\!
\right\}\!\!,\!\!\!
\end{equation}
\begin{equation}
G(X)=\frac{\sin 2\theta M (\epsilon_0-E) K_2(\sqrt{M (\epsilon_0-E)}X/\hbar\sin\theta)}{8\hbar^2\pi^3 X^2},
\end{equation}
with $K_2(z)$ being the exponentially decaying Bessel (Macdonald) function. A detailed derivation is given in Ref.~\cite{marcelis} and is omitted here.
The operators $\hat{L}$ and $\hat{L}'$ conserve angular momentum and, expanding the function $f({\bf R})$ in spherical harmonics, we arrive at a set of uncoupled 
1D integral equations for each of the radial functions $f_l(R)$. Below we present the results for $s$-wave dimer-dimer scattering.

At large distances ($R\gg a$) the reduced wavefunction $\Psi({\bf R},{\bf r})$ takes the form:
\begin{equation}\label{asymptote}
\Psi({\bf R},{\bf r})\approx\Psi({\bf R})\psi_{\bf R}^{-}({\bf r}),
\end{equation}
and one can show that
\begin{equation}\label{Psif}
\Psi({\bf R})\propto f({\bf R});\,\,\,\,\,\, R\gg a.
\end{equation}
Therefore, $f({\bf R})$ can serve as the wavefunction for the dimer-dimer motion at large distances. In particular, it contains the dimer-dimer scattering phase shift.

The dimer-dimer $s$-wave scattering amplitude $a_{dd}$ is determined from the asymptotic behavior of the solution of Eq.~(\ref{final}) at large distances 
for $E=2\epsilon_0$, which should be matched with   
\begin{equation}\label{Psifs}
f_0(R)\propto (1/R-1/a_{dd})
\end{equation}
at $R\gg a\ln \sqrt{M/m}$. In Fig.~11 we compare the resulting $a_{dd}/a$ with that following from Eq.~(\ref{addM}). The results agree quite well even for
moderate values of $M/m$. These results also agree with the calculations based on the exact four-body equation for $M/m<13.6$ \cite{PSSJ} and displayed in Fig.~8, and
with the Monte Carlo results for $M/m<20$ \cite{Blume}.

\begin{figure}[h]
\centerline{\includegraphics[width=0.8\hsize,clip]{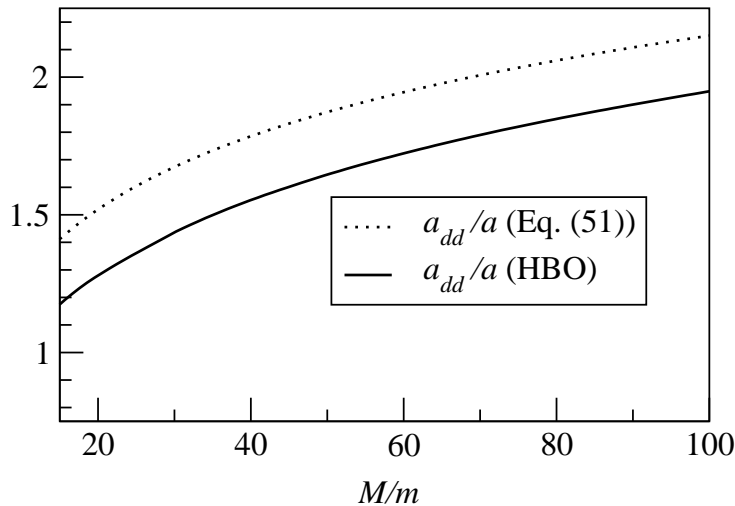}}
\caption{
The dimer-dimer $s$-wave scattering length $a_{dd}/a$. The solid curve shows the results obtained in the hybrid Born-Oppenheimer approximation, and the dotted line 
the results of Eq.~(\ref{addM}). 
\label{ascat}}
\end{figure}

It is straightforward to extend this theory to account for inelastic processes of the trimer formation and the relaxation of dimers into deeply bound states. 
Let us first assume that the rate of the relaxation into deep molecular states is negligible and neglect this process. Then the three-body parameter is real, and the trimer
formation rate
is determined by the imaginary part of the $s$-wave scattering length. The rate constant is given by \cite{LL3} 
\begin{equation}\label{Inelastic}
\alpha = -\frac{16\hbar\pi}{M}\mathrm{Im}(a_{dd}).
\end{equation}
Alternatively, if it is necessary to know the rate of the trimer formation in the state $\nu$, one can substitute the solution of Eq.~(\ref{final}) into
Eq.~(\ref{fourbodysolution}) 
and calculate the flux of light atoms at ${\bf r}\rightarrow \infty$. The summation over $\nu$ gives the same result as Eq.~(\ref{Inelastic}). We find that the contribution of 
the highest ``dangerous'' trimer state is by far dominant and $\alpha$ is very sensitive to its position.

We now include the relaxation of the dimers into deeply bound states. 
As we have mentioned above, the light-heavy-heavy relaxation process can be taken into account by adding an imaginary part to the three-body parameter. 
The total inelastic decay rate is then still given by Eq.~(\ref{Inelastic}). However, strictly speaking we can no longer distinguish between the formation of the trimer in a 
particular state and the collisional relaxation since the trimers ultimately decay due to the relaxation process. In this sense the only decay channel is the relaxation. However,
for a 
sufficiently long lifetime of a trimer, i.e. if the trimer states are narrow resonances, we can still observe a pronounced dependence of the total inelastic decay rate on the 
position of the highest ``dangerous'' trimer state (see below).

Fig.~\ref{YbLi} shows the results for the inelastic collisional rate in the case of bosonic molecules with the mass ratio $M/m=28.5$ characteristic of  
${}^{171}$Yb-${}^6$Li dimers. The solid line corresponds to the case of a real three-body parameter. It is convenient to introduce a related quantity, 
$a_0$, defined as the value of $a$ at which the energy, $\epsilon_\nu$, of a trimer state exactly equals $E=2\epsilon_0$. This new ``dangerous'' trimer state 
becomes more deeply bound for $a>a_0$ and the rate constant rapidly increases. It is proportional to the density of states in the outgoing atom-trimer channel. The corresponding 
orbital angular momentum is equal to 1 and the threshold law reads (see also inset in Fig.~\ref{YbLi}):
\begin{equation}\label{Threshold}
\alpha\propto {\rm const}+(E-\epsilon_\nu)^{3/2}\propto {\rm const}+(a-a_0)^{3/2}.
\end{equation}
The constant term in Eq.~(\ref{Threshold}) describes the contribution of more deeply bound states, which is typically very small. In fact, as trimer states 
become more compact, both light atoms should approach the heavy atoms and each other to small distances where the trimer formation takes place. Since they are identical fermions, 
there is a strong suppression of the trimer formation to these deeply bound states. 

\begin{figure}[h]
\centerline{\includegraphics[width=0.8\hsize,clip]{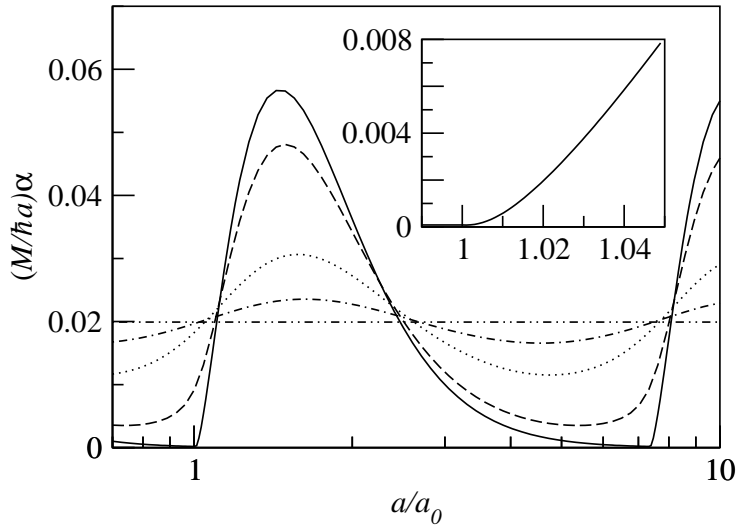}}
\caption{
The inelastic rate constant for bosonic dimers with $M/m=28.5$ as a function of the atom-atom scattering length $a$. The solid line corresponds to the case of a real three-body 
parameter. The results plotted in dashed, dotted, dash-dotted, and dash-dot-dot lines are obtained by taking into account the light-heavy-heavy relaxation processes. The values 
of the elasticity parameter $\eta_*=$0.1, 0.5, 1, and $\infty$, respectively (see text). The quantity $a_0$ is the value of $a$ at which the energy of a trimer state equals 
$E=2\epsilon_0$ and a new inelastic channel opens. The inset shows the region $a\approx a_0$ in greater detail in order to see the threshold behavior (\ref{Threshold}).
\label{YbLi}}
\end{figure}

The dependence of $\alpha$ on $a/a_0$ is periodic on the logarithmic scale, the multiplicative factor being equal to $\lambda_1\approx 7.3$. The dashed, dotted, and dash-dotted
curves are obtained for $\eta_*=$0.1, 0.5, and 1, respectively. The corresponding values of the ratio 
$\Phi_{\rm out}/\Phi_{\rm in}$ are 0.67, 0.14, and 0.02. The horizontal line represents the limiting case of $\eta_*=\infty$ or $\Phi_{\rm out}=0$. This case is universal in 
the sense that physical observables depend only on the masses and the atomic scattering length.

For a very weak light-heavy-heavy relaxation, the dimer-dimer inelastic collision can be viewed as the formation of a trimer (with the rate constant $\alpha$) followed 
by its slow decay due to the relaxation. In this case one can think of detecting the trimers spectroscopically. We note, however, that even for the conditions corresponding 
to the dashed curve in Fig.~\ref{YbLi}, i.e. for $\eta_*$ as small as 0.1, the decay rate of the trimer $\tau^{-1}\approx 0.25|\epsilon_{\nu}|/\hbar$ is rather fast, which will
likely make 
its direct detection difficult. For larger $\eta_*$ it is impossible to separate the formation of trimers from their intrinsic relaxational decay, and $\alpha$ is practically  
the relaxation rate constant. Remarkably, it remains quite sensitive to the positions of trimer states (in this case resonances) for values of $\eta_*$ up 
to 0.5 and even larger. This suggests that measuring the lifetime of a gas of dimers as a function of $a$ may provide important information on three-body observables. Moreover, 
for small $\eta_*$ it may be possible to crate a stable molecular gas in sufficiently broad regions of $a$, where ``dangerous'' trimer states are far from the trimer formation
threshold.

The inelastic rate $\alpha$ for other mass ratios in the range $20<M/m<76$ has also been found \cite{marcelis}. Its dependence on the scattering length has the same form as
depicted in Fig.~\ref{YbLi}. The maximum of the rate constant is well fitted by the formula $\alpha_{\rm max}= 5.8(\hbar a/M)\exp(-0.87\sqrt{M/m})$, and the position of the 
horizontal line
($\eta_*=\infty$) by $\alpha_{\infty}= 1.6(\hbar a/M)\exp(-0.82\sqrt{M/m})$. The multiplicative factor in the log-periodic dependence is given by Eqs.~(\ref{lambdal})
and (\ref{sl}) with $l=1$.

The same method was employed to estimate the formation of the universal trimer state with the orbital angular momentum $l=1$ at mass ratios $M/m>12.7$ but below the critical value
for the onset of the Efimov effect \cite{marcelis}. 
The rate constant increases with $M/m$ and reaches $\alpha=0.2(\hbar a/M)$ close to the critical mass ratio. This corresponds to the imaginary part of
the scattering length ${\rm Im}a_{dd}\approx 4\times 10^{-3}a$, which is by a factor of $300$ smaller than the real part of $a_{dd}$ obtained from four-body calculations
\cite{PSSJ}. Thus, the formation of this state does not change the elastic scattering amplitude $a_{dd}$ shown in Fig.~8.

We can now estimate the collisional rates for $^{171}$Yb-$^{6}$Li dimers. On the basis of the results in 
Fig.~\ref{YbLi} we find that for $a=20$nm the upper bound of the inelastic rate constant is $\alpha_{\rm max}\approx 4\times 10^{-13}$cm$^3/$s. 
The elastic rate constant for a thermal gas with $a\sim 20$nm and $T\sim 100$nK equals
\begin{equation}\label{Elastic}
\alpha_{el}\approx 8\pi |a_{dd}|^2 \sqrt{2T/M}\sim 4\times 10^{-11}{\rm cm}^3/{\rm s}.
\end{equation}
Here we used the calculated $s$-wave dimer-dimer scattering length $a_{dd}\approx 1.4 a$. We see that $\alpha_{el}$ is much larger than $\alpha$ and this inequality 
becomes even more pronounced for larger $a$ due to the scaling relations  $\alpha_{el}\propto a^2$ and $\alpha_{\rm max}\propto a$. Thus, the gas of molecules of heavy
and light fermions is well suited for evaporative cooling towards their Bose-Einstein condensation.

\section{Crystalline molecular phase}

\subsection{Born-Oppenheimer potential in a many-body system of molecules of heavy and light fermions}  

Strong long-distance repulsive interaction between weakly bound molecules 
of light and heavy fermionic atoms has important consequence 
not only for the relaxation process, but
also for macroscopic properties of the molecular system.
In contrast to two-component Fermi gases of atoms in
different internal states, heteronuclear Fermi-Fermi mixtures can form a 
molecular crystalline phase even when the mean interparticle
separation greatly exceeds the size of the molecule:
\begin{equation}   \label{na3}
{\bar R}\gg a.
\end{equation}

Let us consider a mixture of heavy and light fermionic atoms with 
equal concentrations and a large positive scattering length for the interaction
between them, satisfying the inequality $a\gg R_e$. At zero temperature all
atoms will be converted into weakly bound molecules and under the condition (\ref{na3})
the molecular size ($\sim a$) will be much smaller than the mean intermolecular separation. 
Then, using the Born-Oppenheimer approximation and
integrating out the motion of light atoms we are left with a system of
identical (composite) bosons which is described by the Hamiltonian:
\begin{equation}   \label{HO}
\hat H=-\frac{\hbar^2}{2M}\sum_i\Delta_{{\bf R}_i}+\frac{1}{2}\sum_{i\neq
j}U_{\rm eff}(R_{ij}),
\end{equation}
where indices $i$ and $j$ label the bosons, their coordinates are denoted by
${\bf R}_i$ and ${\bf R}_j$, and $R_{ij}=|{\bf R}_i-{\bf R}_j|$ is the
separation
between the $i$-th and $j$-th bosons. 
Assuming that the motion of light fermions is three-dimensional, the effective repulsive potential $U_{\rm eff}$
is given by Eq.~(\ref{Exchange}) and is independent of the mass of the heavy atom $M$. Therefore, at a large mass ratio 
$M/m$ it dominates over the kinetic energy which is inversly proportional to $M$, which may lead to the formation of a crystalline phase.

We will discuss the case where the motion of heavy atoms is confined to two dimensions, while the motion of light atoms can be either
2D or 3D. It will be shown that the Hamiltonian (\ref{HO}) with $U_{\rm eff}$ (\ref{Exchange}), supports the first order quantum 
gas-crystal transition at $T=0$ \cite{PAPSS}. This phase transition resembles the one for the flux lattice melting 
in superconductors, where the flux lines are mapped onto a system of bosons interacting  
via a 2D Yukawa potential \cite{Nelson}. In this case the Monte Carlo studies \cite{Ceperley,Blatter} 
identified the first order liquid-crystal transition at zero and finite temperatures. Aside from the difference 
in the interaction potentials, a distinguished feature of our system is related to its stability. The molecules
can undergo collisional relaxation into deeply bound states, or form weakly bound trimers. Another 
subtle question is how dilute the system should be to enable the use of the binary approximation for the molecule-molecule
interaction, leading to Eqs.~(\ref{HO}) and (\ref{Exchange}). 

Let us first consider the system of $N$ molecules and derive the Born-Oppenheimer interaction potential for this system.  
Omitting the interaction between light (identical) fermions, it is sufficient to 
find $N$ lowest single-particle eigenstates, and the sum of their energies will give the 
interaction potential for the molecules. For the interaction between light and heavy atoms 
we use the Bethe-Peierls approach, and the wavefunction of a single light atom then reads:
\begin{equation}   \label{psi} 
\Psi(\{{\bf R}\},{\bf r})=\sum_{i=1}^{N}C_{i}G_{\kappa}({\bf r}-{\bf R}_i),
\end{equation}
where ${\bf r}$ is its coordinate. The Green function $G_{\kappa}$ satisfies the equation 
$(-\nabla_{\bf r}^2 +\kappa^2)G_{\kappa}({\bf r})=\delta({\bf r})$. The energy of the state (\ref{psi}) equals 
$\epsilon=-\hbar^2\kappa^2/2m$, and here we only search for negative single-particle energies. The dependence of 
the coefficients $C_{i}$ and $\kappa$ on $\{{\bf R}\}$ is obtained using the Bethe-Peierls boundary condition:
\begin{equation}      \label{BP}
\Psi(\{{\bf R}\},{\bf r})\propto G_{\kappa_0}({\bf r}-{\bf R}_i);
\,\,\,\,\,{\bf r}\rightarrow {\bf R}_i.
\end{equation}
Up to a normalization constant, $G_{\kappa_0}$ is the wavefunction of a bound state of a single molecule with 
energy $\epsilon_0=-\hbar^2\kappa_0^2/2m$ and molecular size $\kappa_0^{-1}$. The Bethe-Peierls boundary condition (\ref{BP}) 
written through the Green function $G_{\kappa_0}$ can be used for both 2D and 3D motion of light atoms. In the latter case one has
$\kappa_0=a^{-1}$, the Green function is $G_{\kappa_0}({\bf r}-{\bf R}_i)\propto (|{\bf r}-{\bf R}_i|^{-1}-a^{-1})$ for 
$|{\bf r}-{\bf R}_i|\rightarrow 0$, and Eq.~(\ref{BP}) takes the form (\ref{boundary1}).

From Eqs.~(\ref{psi}) and (\ref{BP})  
we get a set of $N$ equations: $\sum_jA_{ij}C_j=0$, where $A_{ij}=\lambda(\kappa)\delta_{ij}+G_{\kappa}(R_{ij})(1-\delta_{ij})$, 
$R_{ij}=|{\bf R}_i-{\bf R}_j|$, and $\lambda(\kappa)=\lim_{r\rightarrow 0}[G_{\kappa}(r)-G_{\kappa_0}(r)]$. 
The single-particle energy levels are determined by the equation
\begin{equation}     \label{det}
{\rm det}\left[ A_{ij}(\kappa,\{{\bf R}\})\right]=0.
\end{equation}

For $R_{ij}\rightarrow \infty$, Eq.~(\ref{det}) gives an N-fold degenerate 
ground state with $\kappa=\kappa_0$. At finite large $R_{ij}$,  
the levels split into a narrow band. Given a small parameter 
\begin{equation}     \label{ksi}
\xi=G_{\kappa_0}(\tilde R)/\kappa_0 |\lambda'_\kappa(\kappa_0)|\ll 1,
\end{equation} 
where $\tilde R$ is a characteristic distance at which heavy atoms can approach each other, the bandwidth is 
$\Delta\epsilon\approx4|\epsilon_0|\xi\ll |\epsilon_0|$. 
It is important for the adiabatic approximation that
all lowest $N$ eigenstates have negative energies and are separated from the continuum by a gap $\sim |\epsilon_0|$. 

We now calculate the single-particle energies up to second order in $\xi$. To this order we write
$\kappa(\lambda)\approx \kappa_0+\kappa'_\lambda\lambda+\kappa''_{\lambda\lambda}\lambda^2/2$ and turn from 
$A_{ij}(\kappa)$ to $A_{ij}(\lambda)$:
\begin{equation}       \label{Aij}
A_{ij}=\lambda\delta_{ij}+[G_{\kappa_0}(R_{ij})+\kappa'_\lambda\lambda 
\partial G_{\kappa_0}(R_{ij})/\partial\kappa](1-\delta_{ij}),
\end{equation}
where all derivatives are taken at $\lambda=0$. Using $A_{ij}$~(\ref{Aij}) in Eq.~(\ref{det}) 
gives a polynomial of degree $N$ in $\lambda$. Its roots $\lambda_i$ give the light-atom energy spectrum 
$\epsilon_i=-\hbar^2\kappa^2(\lambda_i)/2m$. The total energy, $E=\sum_{i=1}^{N}\epsilon_i$, is then given by 
\begin{equation}  \label{E}
E=-(\hbar^2/2m)\Big[ N\kappa_0^2+2\kappa_0\kappa'_\lambda\sum_{i=1}^{N}\lambda_i 
+(\kappa\kappa'_\lambda)'_\lambda\sum_{i=1}^{N}\lambda_i^2\Big].
\end{equation}
Keeping only the terms up to second order in $\xi$ and using 
basic properties of determinants and polynomial roots we find that the first order terms vanish, and 
the energy reads $E=N\epsilon_0+(1/2)\sum_{i\neq j}U(R_{ij})$, where 
\begin{equation}    \label{U}
U(R)=-\frac{\hbar^2}{m}\Big[\kappa_0(\kappa'_\lambda)^2
\frac{\partial G_{\kappa_0}^2(R)}{\partial\kappa}+(\kappa\kappa'_\lambda)'_\lambda G_{\kappa_0}^2(R)\Big].
\end{equation}
Thus, up to second order in $\xi$ the interaction in the system of N molecules is the sum of binary 
potentials (\ref{U}). 

If the motion of light atoms is 3D, the Green function is 
$G_{\kappa}(R)=(1/4\pi R)\exp(-\kappa R)$, and $\lambda(\kappa)=(\kappa_0-\kappa)/4\pi$,
with the molecular size $\kappa_0^{-1}$ equal to the 3D scattering length $a$.
Equation (\ref{U}) then gives a repulsive potential $U_{\rm ex}$ (\ref{Exchange}) which we now denote as
$U_{\rm ex}^{3D}$:
\begin{equation}     \label{Yukawa}
U_{\rm ex}^{3D}(R)=4|\epsilon_0|( 1-(2\kappa_0R)^{-1})\exp(-2\kappa_0R)/\kappa_0R,
\end{equation}
and the criterion (\ref{ksi}) reads $(1/\kappa_0R)\exp(-\kappa_0R)\ll 1$. 
For the 2D motion of light atoms we have $G_{\kappa}(R)=(1/2\pi)K_0(\kappa R)$ and 
$\lambda(\kappa)=-(1/2\pi)\ln(\kappa/\kappa_0)$, where $K_0$ is the decaying Bessel function,
and $\kappa_0^{-1}$ follows from \cite{PS2001}. 
This leads to a repulsive intermolecular potential:
\begin{equation}     \label{U2D}
U_{\rm ex}^{2D}(R)=4|\epsilon_0|[\kappa_0RK_0(\kappa_0R)K_1(\kappa_0R)-K_0^2(\kappa_0R)],
\end{equation}
with the validity criterion $K_0(\kappa_0R)\ll 1$. In both cases, which we denote $2\times 3$ and $2\times 2$ 
for brevity, the validity criteria are well satisfied already for $\kappa_0 R\approx 2$.

\subsection{Gas-crystal quantum transition}

The inequality $\kappa_0{\bar R}\gtrsim 2$ may be considered as the condition under which the system is
described by the Hamiltonian (\ref{HO}), with $U_{eff}$ given by Eq.~(\ref{Yukawa}) for the 3D motion of light atoms,
or by Eq.~(\ref{U2D}) in the case where this motion is 2D. The state of the system is then 
determined by two parameters: the mass ratio $M/m$ and
the rescaled 2D density $n\kappa_0^{-2}$. At a large $M/m$, the potential repulsion dominates over
the kinetic energy, which should lead to the formation of a crystalline ground state. For separations $R_{ij}<\kappa_0^{-1}$ the adiabatic approximation breaks down. 
However, the interaction potential $U(R)$ is strongly repulsive at larger distances. 
Hence, even for an average separation between heavy atoms, ${\bar R}$, close to $2/\kappa_0$, 
they approach each other to distances smaller than $\kappa_0^{-1}$ with a small tunneling probability  
$P\propto\exp (-\beta \sqrt{M/m})\ll 1$, where $\beta\sim 1$. It is then possible to extend $U_{\rm ex}^{3D}(R)$ (or $U_{\rm ex}^{2D}(R)$) to 
$R\lesssim \kappa_0^{-1}$ in a way providing a proper 
molecule-molecule scattering phase shift in vacuum and verify that the phase diagram 
for the many-body system is not sensitive to the choice of this extension \cite{PAPSS}. 

In Fig.~13 we display the zero-temperature phase diagram obtained by
the Diffusion Monte Carlo method \cite{PAPSS}. Simulations were performed
with $30$ particles and showed that the solid phase is a 2D triangular lattice. 
For the largest density it has been verified that using more particles has little effect on the results.

\begin{figure}[h]
\label{PhaseDiagram}
\centerline{\includegraphics[width=0.8\hsize]{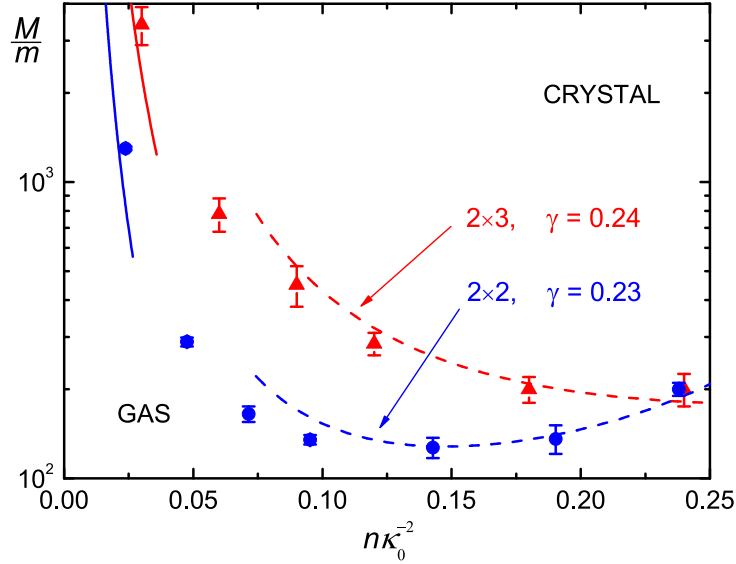}}
\caption{
Diffusion Monte Carlo gas-crystal transition lines for 3D (triangles) and 2D (circles) 
motion of light atoms. Solid curves show the low-density hard-disk limit, 
and dashed curves the results of the harmonic approach (see text).
}
\end{figure}

For both $2\times 3$ and $2\times 2$ cases the (Lindemann) ratio $\gamma$ of the rms displacement of molecules to  
${\bar R}$ on the transition lines ranges from $0.23$ to $0.27$.  
At low densities $n$ the de Broglie wavelength of molecules is 
$\Lambda\sim\gamma {\bar R}\gg\kappa_0^{-1}$, and $U_{\rm ex}^{3D}(R)$ (or $U_{\rm ex}^{2D}(R)$) can be approximated by a hard-disk potential with the diameter
equal to the 2D scattering length. 
Then, using the DMC results for hard-disk bosons \cite{Xing}, we obtain the transition 
lines shown by solid curves in Fig.~13. At larger $n$, we have 
$\Lambda<\kappa_0^{-1}$ and using the
harmonic expansion of $U(R)$ around equlibrium positions in the crystal,
we calculate the Lindemann ratio and select $\gamma$ for the best
fit to the Monte Carlo data points (dashed curves in Fig.~13).

\subsection{Molecular superlattice in an optical lattice}

The mass ratio above 100, required for the observation of the crystalline order can be achieved 
in an optical lattice with a small filling factor for heavy atoms. Their effective mass in the lattice, $M_*$,
can be made very large, and the discussed solid phase  
should appear as a superlattice. 
There is no interplay between the superlattice order and the shape of the underlying optical lattice,
in contrast to the recently studied solid and supersolid phases in a triangular lattice with the filling 
factor of order one \cite{wessel,heidarian,melko}.
The superlattice discussed in our review remains compressible and supports two branches of phonons. 

The gaseous and solid phases of weakly bound molecules in an optical lattice are metastable. As well as in
the gas of such molecules in free space, the main decay channels are the relaxation of molecules into deep bound states 
and the formation of trimer states by one light and two heavy atoms. The relaxation into
deeply bound states turns out to be rather slow, with a relaxation time exceeding 10 s even at
2D densities $\sim 10^9$ cm$^{-2}$ \cite{PAPSS}.  
 
The most interesting is the formation of the trimer states.
In an optical lattice the trimers are eigenstates of the Hamiltonian 
$H_0=-(\hbar^2/2M_*)\sum_{i=1,2}\Delta_{\bf R_i}+\epsilon_{+}(R_{12})$. In a deep lattice it is possible to 
neglect all higher bands and regard ${\bf R_i}$ as discrete lattice coordinates and $\Delta$ as the lattice Laplacian. 
Then, the fermionic nature of the heavy atoms prohibits them to be on the same lattice site. For a very 
large mass ratio $M_*/m$ the kinetic energy term in $H_0$ can be neglected, and the lowest trimer state has energy 
$\epsilon_{\rm tr}\approx \epsilon_{+}(L)$, where $L$ is the lattice period. It consists of a pair of heavy atoms 
localized at neighboring sites and a light atom in the gerade state. Higher trimer states are formed by 
heavy atoms localized in sites separated 
by distances $R>L$. This picture breaks down at large $R$, where the spacing between trimer levels
is comparable with the tunneling energy $\hbar^2/M_*L^2$ and the heavy atoms are delocalized. 

In the many-body molecular system the scale of energies in Eq.~(\ref{HO}) is much smaller than $|\epsilon_0|$. Thus, 
the formation of trimers in molecule-molecule ``collisions'' is energetically allowed only if the trimer binding energy 
is $\epsilon_{\rm tr}<2\epsilon_0$. Since the lowest trimer energy in the optical lattice is $\epsilon_{+}(L)$, the trimer formation
requires the condition $\epsilon_{+}(L)\lesssim 2\epsilon_0$, which is equivalent to $\kappa_0^{-1}\gtrsim 1.6 L$ in the 
$2\times 3$ case and $\kappa_0^{-1}\gtrsim 1.25 L$ in the $2\times 2$ case. This means that for a sufficiently small 
molecular size or large lattice period $L$ the formation of trimers is forbidden. 

At a larger molecular size or smaller $L$ the trimer formation is possible. The formation rate has been calculated
in Ref.~\cite{PAPSS} by using the hybrid Born-Oppenheimer approach, and here we only present the results and give 
their qualitative explanation. In order to form a bound trimer state two molecules have to tunnel towards each other at distances $R\lesssim \kappa_0^{-1}$.
This can be viewed as tunneling of particles with mass $M_*$ in the repulsive potential $U_{\rm eff}(R)$. Therefore, the
probability of approaching at interheavy separations where the trimer formation occurs, acquires a small factor
$\exp(-J\sqrt{M_*/m})$ with $J\sim 1$, and so does the formation rate. Thus, one can suppress the trimer formation by increasing the ratio 
$M_*/m$. On the other hand, for $M_*/m\lesssim 100$ these peculiar bound states can be formed on the time scale 
$\tau\lesssim 1$ s. 

Note that the trimer states in an optical lattice, at least the lowest ones,  are much more long-lived than in the gas phase. 
An intrinsic relaxational decay is strongly suppressed as it requires the two heavy atoms of the trimer
to approach each other and occupy the same lattice sites. The trimer state can also decay when one of the heavy
atoms of the trimer is approached by its own light atom and another light atom at light-heavy separations $\sim R_e$.
However, this decay channel turns out to be rather slow, with a decay time exceeding tens of seconds even at 2D densities
$n\sim 10^9$ cm$^{-2}$ \cite{PAPSS}.

\section{Concluding remarks and prospects}  

The most distinguishing feature of weakly bound bosonic molecules formed of fermionic
atoms is their remarkable collisional stability, despite they 
are in the highest ro-vibrational state. As we mentioned in the Introduction, 
the lifetime of such molecules can be of the order of seconds or even tens of seconds at densities
$\sim 10^{13}$ cm$^{-3}$, depending on the value of the two-body scattering length.
This allows for interesting manipulations with these molecules. One of the ideas is 
related to reaching extremely low temperatures in a gas of fermionic atoms at 
$a<0$ and achieving the superfluid BCS regime. This regime has not been obtained 
so far because of difficulties with evaporative cooling of fermionic atoms due to Pauli 
blocking of their elastic collisions. The route to BCS may be the following. 
In the first stage, one arranges a deep evaporative cooling of the molecular
Bose-condensed gas to temperatures of the order of the chemical
potential. Then one converts the molecular BEC into fermionic
atoms by adiabatically changing the scattering length to negative
values. This provides an additional cooling, and the obtained atomic
Fermi gas will have extremely low temperatures $\lesssim 10^{-2}T_F$,
where $T_F$ is the Fermi temperature. The gas can then enter
the superfluid BCS regime \cite{carr}. Moreover, at such temperatures
elastic collisions are suppressed by a very strong Pauli blocking and
the thermal cloud is in the collisionless regime. This is
promising for identifying the BCS-paired state through the
observation of collective oscillations or free expansion
\cite{stringari,O'Hara,ens2,rudy3,Kinast}.

It will also be interesting to transfer weakly bound molecules
of fermionic atoms to their ground (or less excited) ro-vibrational state.
For molecules of bosonic atoms this has been done using
two-photon spectroscopy \cite{DeM,DeM2,Rudilast} and by magnetically tuned 
mixing of neighbouring molecular levels, which enables otherwise forbidden 
radiofrequency transitions \cite{Lang}. Long lifetimes of weakly bound
molecules of fermionic atoms at densities $\sim 10^{13}$ cm$^{-3}$
may ensure an efficient production of ground state
molecules compared to the case of more short-lived molecules of bosonic atoms.
One could then extensively study the physics of molecular
Bose-Einstein condensation. Moreover, the ground-state heteronuclear molecules
have a relatively large permanent dipole moment and can be polarized by an electric field. This may be used to create a gas
of dipoles interacting via anisotropic long-range forces, which
drastically changes the physics of Bose-Einstein condensation
(see, e.g., \cite{dipoles} and references therein).

In the last few years, the observation of the Efimov effect was one of the important
goals in the cold atom studies. As we discussed in Section 3, the Efimov trimers 
in the gas phase are short-lived and rather represent narrow resonances.
The Efimov effect then manifests itself in the log-periodic dependence of collision rates on the two-body
scattering length. In particular, this is the case for the rate of three-body recombination of atoms \cite{kraemer06} 
and for the rate of trimer formation in molecule-molecule collisions \cite{marcelis}.
In this sense, the trimer formation in gases of bosonic molecules consisting of heavy and light fermions (such as, for example, LiYb)
attracts a great interest as the observation of the Efimov oscillations requires a much smaller change of the 
two-body scattering length (by a factor of 7 or 5) than in the case of identical bosons.

Of particular interest are the trimer states of two heavy and one light fermion in an optical lattice.
For 2D densities $\sim 10^8$ cm$^{-2}$ the rate of the trimer formation can be of the order of seconds, and these 
states can be detected optically. As we already mentioned in Section 4, the lattice trimers are long-lived, 
with a lifetime that can be of the order of tens of seconds. Thus, it is interesting to study to which extent
these non-conventional states, in which the heavy atoms are localized in different sites and the light atom is 
delocalized in between them, can exhibit the Efimov effect.

The creation of a superlattice of molecules in an optical lattice also looks feasible. 
A promising candidate is the $^6$Li-$^{40}$K mixture as the Li atom may tunnel freely in a lattice while localizing 
the heavy K atoms to reach high mass ratios. A lattice with 
period 250 nm and K effective mass M*= 20M provide a tunneling 
rate $\sim 10^3$ s$^{-1}$ sufficiently fast to let the crystal form. Near 
a Feshbach resonance, a value $a=500$ nm gives 
a binding energy $300$ nK, and lower temperatures  
should be reached in the gas. The parameters $n\kappa_0^{-2}$ of Fig.~13 are then 
obtained at 2D densities in the range $10^7-10^8$ cm$^{-2}$ easily reachable in 
experiments. 

\subsection*{Acknowledgements}

The work on this review was financially supported by the IFRAF Institute, by
ANR (grants 05-BLAN-0205 and and 06-Nano-014), by the EuroQUAM program of ESF
(project Fermix), by Nederlandse Stichtung voor Fundamenteel 
Onderzoek der Materie (FOM), and by the Russian Foundation for Fundamental
Research. LKB is a research unit no. 8552 of CNRS, ENS, and of the University
of Pierre et
Marie Curie. LPTMS is a mixed research unit no. 8626 of CNRS and University
Paris-Sud. 



\begin{thebibliography}{99}
\bibitem{comment1} Including $p$-wave interactions, one can think of the
formation of deeply bound states by two identical ($\uparrow$ or $\downarrow$) fermions.
So, the collision of two weakly bound molecules can lead to the creation of a
deep bound state by two $\uparrow$ (or $\downarrow$) fermionic atoms, and two
$\downarrow$ (or $\uparrow$) atoms become unbound.
\end{thebibliography}
\end{document}